\documentclass[reprint,nofootinbib,amsmath,longbibliography,amssymb,aps,pra,a4paper,notitlepage,superscriptaddress]{revtex4-1}

\usepackage[latin1]{inputenc}
\usepackage[english]{babel}
\usepackage{graphicx}% Include figure files
\usepackage{dcolumn}% Align table columns on decimal point
\usepackage{natbib}
\usepackage{hyperref}
\usepackage{multirow}
\usepackage{bm}% bold math

\usepackage{changes}
\usepackage{soul}

\usepackage{bibunits}

\usepackage{letltxmacro}

\LetLtxMacro{\ORIGselectlanguage}{\selectlanguage}
\makeatletter
\DeclareRobustCommand{\selectlanguage}[1]{%
  \@ifundefined{alias@\string#1}
    {\ORIGselectlanguage{#1}}
    {\begingroup\edef\x{\endgroup
       \noexpand\ORIGselectlanguage{\@nameuse{alias@#1}}}\x}%
}
\newcommand{\definelanguagealias}[2]{%
  \@namedef{alias@#1}{#2}%
}
\makeatother

\definelanguagealias{en}{english}

\newcommand{\roo}{\rho_{\mathrm{00}}}
\newcommand{\roe}{\rho_{\mathrm{01}}}
\newcommand{\reo}{\rho_{\mathrm{10}}}
\newcommand{\ree}{\rho_{\mathrm{11}}}

\newcommand{\droo}{\dot{\rho}_{\mathrm{00}}}
\newcommand{\droe}{\dot{\rho}_{\mathrm{01}}}
\newcommand{\dreo}{\dot{\rho}_{\mathrm{10}}}
\newcommand{\dree}{\dot{\rho}_{\mathrm{11}}}

\usepackage{xcolor}
\usepackage{xparse,xcoffins}

\ExplSyntaxOn
\NewCoffin\imagecoffin
\NewCoffin\labelcoffin

\keys_define:nn { miguel/label }
{
label   .tl_set:N = \l_miguel_label_tl,
labelbox .bool_set:N = \l_miguel_label_box_bool,
labelbox .default:n = true,
fontsize .tl_set:N = \l_miguel_label_size_tl,
fontsize .initial:n = \footnotesize,
pos .choice:,
pos/nw .code:n = \tl_set:Nn \l_miguel_label_pos_tl { left,up },
pos/ne .code:n = \tl_set:Nn \l_miguel_label_pos_tl { right,up },
pos/sw .code:n = \tl_set:Nn \l_miguel_label_pos_tl { left,down },
pos/se .code:n = \tl_set:Nn \l_miguel_label_pos_tl { right,down },
pos/n .code:n = \tl_set:Nn \l_miguel_label_pos_tl { hc,up },
pos/w .code:n = \tl_set:Nn \l_miguel_label_pos_tl { left,vc },
pos/s .code:n = \tl_set:Nn \l_miguel_label_pos_tl { hc,down },
pos/e .code:n = \tl_set:Nn \l_miguel_label_pos_tl { right,vc },
pos .initial:n = nw,
unknown .code:n   = \clist_put_right:Nx \l_miguel_label_clist
                       ~ { \l_keys_key_tl = \exp_not:n { #1 } }
}

\clist_new:N \l_miguel_label_clist
\box_new:N \l_miguel_label_box
\box_new:N \l_miguel_label_image_box

\NewDocumentCommand{\xincludegraphics}{O{}m}
{
\group_begin:
\tl_clear:N \l_miguel_label_tl
\clist_clear:N \l_miguel_label_clist
\keys_set:nn { miguel/label } { #1 }
\tl_if_empty:NTF \l_miguel_label_tl
	{\miguel_includegraphics:Vn \l_miguel_label_clist { #2 }}
	{
	\SetHorizontalCoffin\imagecoffin
    		{\miguel_includegraphics:Vn \l_miguel_label_clist { #2 }}
    		\SetHorizontalCoffin\labelcoffin
    		{
    		\raisebox{\depth}
       		{
       		\bool_if:NTF \l_miguel_label_box_bool
         	{\fcolorbox{white}{white}{\l_miguel_label_size_tl\l_miguel_label_tl}}
         	{\l_miguel_label_size_tl\l_miguel_label_tl}
       		}
    		}
   \SetVerticalPole\imagecoffin{left}{3pt+\CoffinWidth\labelcoffin/2}
   \SetVerticalPole\imagecoffin{right}{\Width-3pt-\CoffinWidth\labelcoffin/2}
   \SetHorizontalPole\imagecoffin{up}{\Height-4pt-\CoffinHeight\labelcoffin/2}
   \SetHorizontalPole\imagecoffin{down}{3pt+\CoffinHeight\labelcoffin/2}
   \use:x{\JoinCoffins\imagecoffin[\l_miguel_label_pos_tl]\labelcoffin[vc,hc]} 
   \TypesetCoffin\imagecoffin
   }
\group_end:
}

\cs_new_protected:Nn \miguel_includegraphics:nn
{\includegraphics[#1]{#2}}
\cs_generate_variant:Nn \miguel_includegraphics:nn { V }

\ExplSyntaxOff

\begin{document}

\title{Molecular platform for frequency upconversion at the single-photon level}

\author{Philippe Roelli}
\affiliation{EPFL, Swiss Federal Institute of Technology, Institute of Physics, Lausanne, Switzerland}
\author{Diego Martin-Cano}
\affiliation{Max Planck Institute for the Science of Light, Erlangen, Germany}
\author{Tobias J. Kippenberg}
\email[corresponding authors: ]{tobias.kippenberg@epfl.ch}
\affiliation{EPFL, Swiss Federal Institute of Technology, Institute of Physics, Lausanne, Switzerland}
\author{Christophe Galland}
\email{chris.galland@epfl.ch}
\affiliation{EPFL, Swiss Federal Institute of Technology, Institute of Physics, Lausanne, Switzerland}

\date{\today}

\begin{abstract}
Direct detection of single photons at wavelengths beyond 2 microns under ambient conditions
remains an outstanding technological challenge.
One promising approach is frequency upconversion into the visible (VIS) or near-infrared (NIR) domain, 
where single photon detectors are readily available. 
Here, we propose a nanoscale solution based on a molecular optomechanical 
platform to up-convert photons 
from the far and mid-infrared (covering part of the THz gap)
into the VIS-NIR domain. 
{We perform a detailed analysis of its outgoing noise spectral density 
and conversion efficiency with a full quantum model.}
Our platform consists in {doubly resonant nanoantennas focusing 
both the incoming long-wavelength radiation 
and the short-wavelength pump laser field into the same active region. 
There, infrared active vibrational modes are resonantly excited and couple 
through their Raman polarizability to the pump field. 
This optomechanical interaction is enhanced by the antenna and leads to the 
coherent transfer of the incoming low-frequency signal onto the anti-Stokes sideband of the pump laser.}
Our calculations demonstrate that our scheme is realizable with current technology 
and that optimized platforms can reach single photon sensitivity
in a spectral region where this capability remains unavailable to date.
\end{abstract}

\maketitle

%%%%%%%%%%%%%%%%%%%%%%%%%%%%%%%%%%%%%%%%%%%%%%%%%%%%%%%%%%%%%%%%%%%%%%%%%%%%%%%%%%%%%%%%%%%%%%%%%%%%%%%%%%%%%%%%%%%%%%%%%%%%%%%%%%%%%%%%%%%%%%%%%%%%%%%%%%%%%%%%%%%%%%%%%%%%%%%%
%\tableofcontents
\section{Introduction}
Many applications in security, material science and healthcare would benefit from the development 
of new technologies for far and mid-infrared (FIR and MIR, respectively) detection and thermal imaging \cite{tonouchi_cutting-edge_2007}. 
Driven by applications in astronomy, novel cryogenic detectors in the FIR range appeared in the past few years 
\cite{ariyoshi_terahertz_2016,bueno_full_2017}. 
However the ability to efficiently manipulate such electromagnetic signals at room temperature 
is still lacking \cite{sizov_terahertz_2018,rogalski_two-dimensional_2019}. 
In particular, single photon detection, which is now routine in the VIS-NIR region (wavelength in vacuum from 400 to 2000~nm), 
remains impossible or unpractical at longer wavelengths.  
The development of new detection devices operating without complex cryogenic apparatus, 
and featuring improved sensitivity, lower noise and reduced footprint,  
would significantly impact sensing, imaging, spectroscopy and communication technologies. 

In this work we propose a new route to achieve low-noise detection of non-coherent radiation 
between 5-50 THz 
by leveraging optomechanical transduction with molecules \cite{roelli_molecular_2016}, 
whose natural oscillation frequencies are resonant with the incoming field. 
Our strategy consists in converting the incoming low-frequency signal 
onto the anti-Stokes sideband of a pump laser in the VIS-NIR domain, 
where detectors with single photon sensitivity are readily available 
\cite{goltsman_picosecond_2001,hadfield_single-photon_2009}. 
This approach is inspired by the recent realization of coherent frequency 
conversion using different types of optomechanical cavities 
\cite{tian_optical_2010,dong_optomechanical_2012,hill_coherent_2012,
palomaki_coherent_2013,bochmann_nanomechanical_2013,andrews_bidirectional_2014,
forsch_microwave--optics_2020} 
and is conceptually distinct from a recently demonstrated detection scheme 
assisted by a microfabricated resonator \cite{belacel_optomechanical_2017}. 
As an outlook, we propose to leverage constructive interference 
between signals coming from an array of coherently pumped up-converters 
in order to increase further the strength of the converted signal over the incoherent thermal noise. 

While coherent conversion from the MIR to the VIS-NIR domain has so far 
 been achieved by sum-frequency generation in bulk nonlinear crystals
\cite{boyd2008nonlinear,karstad_detection_2005, 
tidemand-lichtenberg_mid-infrared_2016, tseng_upconversion_2018}, 
these schemes operate under several watts of pump power 
and require phase-matching between the different fields propagating in the crystal. 
Our scheme, on the contrary, relies solely on the spatial overlap of the two incoming fields. 
Indeed, we use a nanometer-scale dual antenna that confines 
both electromagnetic fields into similar sub-wavelength mode volumes. 
The optomechanical interaction with the vibrational system takes place in the near field, 
without need to fulfill a phase matching condition. 
Moreover, thanks to the giant field enhancement provided by plasmonic nanogaps, 
the required pump power to achieve efficient conversion is dramatically reduced.

The protocol that we introduce leverages the intrinsic ability of specific molecular vibrations 
to interact both resonantly with MIR-FIR fields and parametrically with VIS-NIR fields, 
as routinely observed in infrared absorption and Raman spectroscopy, respectively. 
The wealth of accessible vibrational modes and frequencies 
\cite{herzberg_molecular_1960,colthup_introduction_1990} 
offers a convenient toolbox to realize efficient frequency upconversion 
in the technologically appealing region of thermal imaging. 

We first introduce the framework describing the interaction between a molecular vibration 
and two electromagnetic fields, one that is resonant with the vibrational frequency, 
the other one that is parametrically coupled to it through the molecular polarization. 
We compute the conversion efficiency and the noise figures of merit of our novel device 
as a function of the optical pump detuning and power. 
We illustrate the achievable performance with a device operating at 30~THz (10~$\mu$m) and find internal conversion efficiencies on the order 
of a few percent and noise-equivalent power below $10^{-12}$ W/$\sqrt{\mathrm{Hz}}$. 
Finally, we demonstrate how our approach may be used to 
reach single-photon detection at frequencies down to $\sim5$~THz. 

%%%%%%%%%%%%%%%%%%%%%%%%%%%%%%%%%%%%%%%%%%%%%%%%%%%%%%%%%%%%%%%%%%%%%%%%%%%%%%%%%%%%%%%%%%%%%%%%%%%%%%%%%%%%%%%%%%%%%%%%%%%%%%%%%%%%%%%%%%%%%%%%%%%%%%%%%%%%%%%%%%%%%%%%%%%%%%%%
\section{Optical conversion scheme}
We start with the description of 
the two types of interactions 
leveraged in the conversion process and describe the relevant parameters. 
For simplicity, we now use the abbreviation IR to denote MIR or FIR fields, 
depending on the vibrational frequency considered.  
First we model the resonant absorption process. 
We assume that the vibrational system is weakly driven, 
meaning that the average number of excited collective vibrational quanta 
is much smaller that the total number of molecular oscillators
coupled to the incoming field, $N_{\textsc{ir}}$. 
At the single-molecule level, this easily satisfied condition corresponds to neglecting transitions beyond the ground and first excited vibrational states.
The collective excitation of an ensemble of vibrational modes can thus be treated as an ensemble of two-level systems \cite{shalabney_coherent_2015}. 

The interaction part of the Hamiltonian is correspondingly approximated by :
\begin{equation}
 \hat{H}_{\mathrm{int}}=
-i\hbar g_{\textsc{ir},0}^{(N_{\textsc{ir}})}\sqrt{\bar{n}_{\textsc{ir}}}\left(\hat{a}_{\textsc{ir}}^{\dagger}
\hat{\sigma}^-_{\nu}+\hat{a}_{\textsc{ir}}\hat{\sigma}^+_{\nu}\right),   
\end{equation}
with $\hat{a}^{\dagger}_{\textsc{ir}},\hat{a}_{\textsc{ir}}$ 
the IR field bosonic ladder operators
and $\hat{\sigma}^+_{\nu},\ \hat{\sigma}^-_{\nu}$ the raising and lowering operators 
of the collective two level system described by a transition frequency 
$\omega_{\nu}$. $g_{\textsc{ir},0}^{(N_{\textsc{ir}})}=\sqrt{N_{\textsc{ir}}}g_{\textsc{ir},0}$ %_{\textsc{ir}}
is the collective resonant vacuum coupling rate of the vibrational mode $\nu$ for $N_{\textsc{ir}}$ identical molecules,
and $\bar{n}_{\textsc{ir}}$ the mean occupation of the IR antenna mode.

The incoming IR field at frequency $\omega_{\textsc{ir}}$ is enhanced by a frequency-matched antenna 
and performs work on the collective transition dipole $\vec{d}_{\nu}$ of the molecular vibration
\cite{cohen-tannoudji_atom-photon_1992}. 
On resonance $\left(\omega_{\textsc{ir}}= \omega_{\nu}\right)$ 
the average number of created phonons is (see Appendix \ref{sec:abs} for a detailed derivation) 
\begin{equation}
\bar{n}_b^{\textsc{ir}}=
\left(\frac{2g_{\textsc{ir},0}^{(N_{\textsc{ir}})}}{\Gamma_{\mathrm{tot}}}\right)^2
\frac{\eta_{\textsc{ir}}}{\kappa^{\textsc{ir}}}|\langle\hat{a}_{\textsc{ir}}^{\mathrm{in}}\rangle|^2 
\label{eq:abs}
\end{equation}
with $|\langle\hat{a}_{\textsc{ir}}^{\mathrm{in}}\rangle|^2$ the incoming IR photon flux. 
In this expression $\kappa^{\textsc{ir}}=\kappa^{\textsc{ir}}_{\mathrm{ex}}+\kappa^{\textsc{ir}}_{0}$ 
is the loss rate of the antenna at the incoming frequency, 
which is the sum of the external decay rate $\kappa^{\textsc{ir}}_{\mathrm{ex}}$ 
(by radiative coupling to the far-field modes) and the internal decay rate $\kappa^{\textsc{ir}}_{0}$ 
(by absorption in the metal). 
$\eta_{\textsc{ir}}=\kappa^{\textsc{ir}}_{\mathrm{ex}}/\kappa^{\textsc{ir}}$ 
is the coupling ratio of the antenna 
and $\Gamma_{\mathrm{tot}}$ the total vibrational decay rate, 
where the intrinsic vibrational linewidth $\Gamma_{\nu}$ is modified 
by its coupling to the IR antenna \cite{metzger_purcell-enhanced_2019} and the optomechanical interaction with the pump laser, as explained below.

\begin{figure}[h!]
	\centering
	\xincludegraphics[width=\columnwidth,label=\textbf{(a)},fontsize=\large]{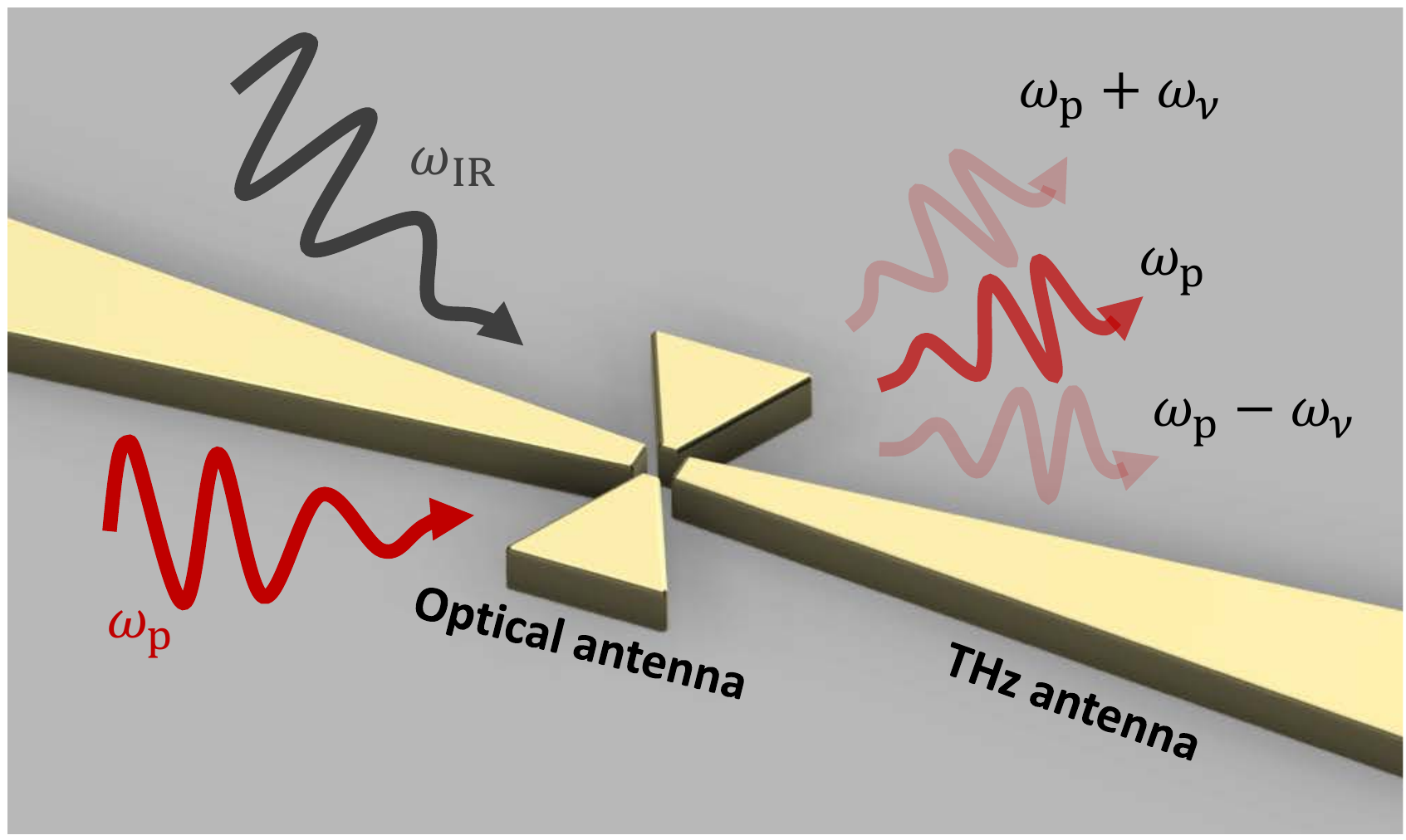}
	\xincludegraphics[width=\columnwidth,label=\textbf{(b)},fontsize=\large]{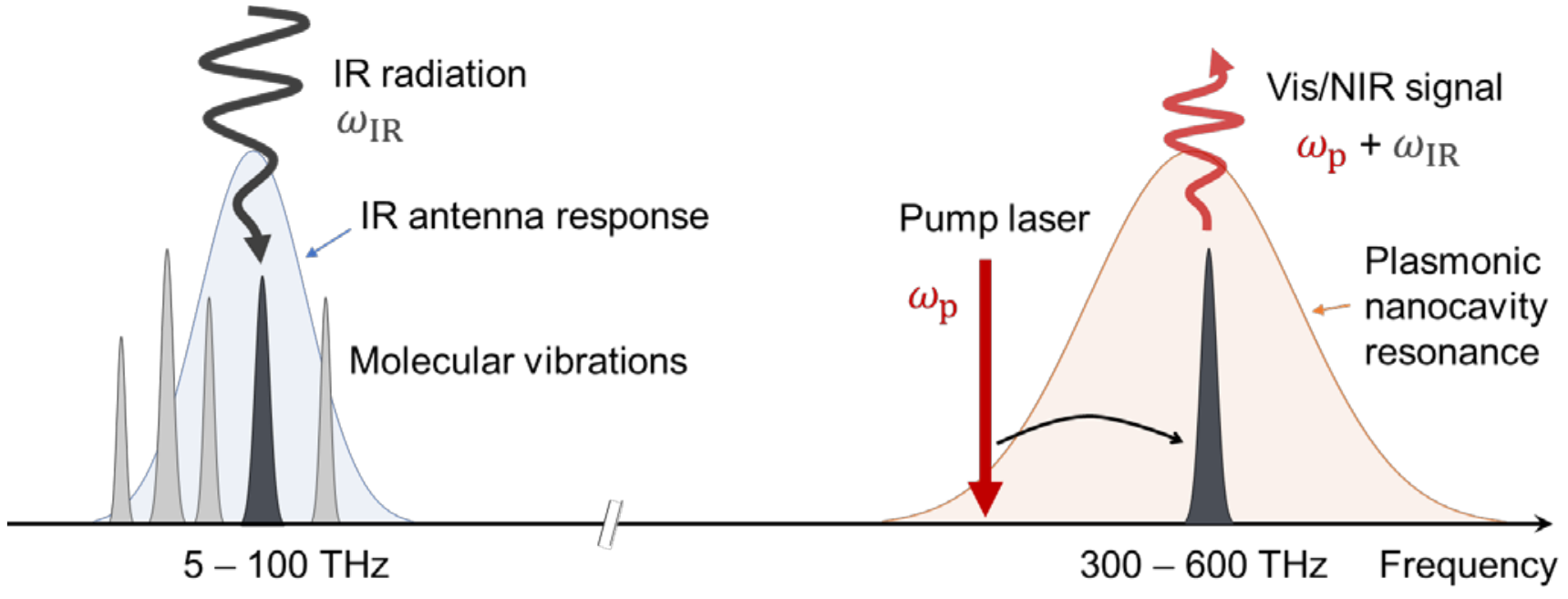}
	\vspace{-20pt}	
	\caption{ \textbf{(a)} Illustration of the envisioned up-conversion device. Both electromagnetic modes 
	are collected with the help of the dual antenna and 
	confined within a volume where molecules are located. 
	\textbf{(b)} Frequency picture of the optomechanical conversion mechanism 
	involving both IR absorption and Raman scattering by specific vibrational modes. 
	Here the pump tone ($\omega_p$) is red-detuned from the optical resonance ($\omega_c\simeq \omega_p+\omega_{\textsc{ir}}$) 
	while the incoming IR signal is resonant with a specific vibrational mode ($\omega_{\textsc{ir}}=\omega_{\nu}$).}
	\label{FreqPic}
	\vspace{-10pt}
\end{figure}
 
As pictured in Fig.~\ref{FreqPic}, we employ a second antenna resonant at $\omega_c$ 
(a frequency in the VIS-NIR domain, which we call ``optical'' domain from here on for brevity),
whose decay rates $\kappa^{\mathrm{opt}}_{\mathrm{ex}},\ \kappa^{\mathrm{opt}}_{0}$ are
defined in the same way as the IR antenna parameters. 
The optical antenna enhances the parametric optomechanical interaction 
of the molecular vibration with a pump laser in the optical domain, as described in Ref. \cite{roelli_molecular_2016}. 
Concisely the interaction between an optical field and $N_{\mathrm{opt}}$ molecular oscillators leads to a dispersive interaction 
described by the Hamiltonian $\hat{H}_{\mathrm{int}}=
-\hbar g_{\mathrm{opt},0}^{(N_{\mathrm{opt}})}\hat{a}_{\mathrm{opt}}^{\dagger}\hat{a}_{\mathrm{opt}}
\left(\hat{b}_{\nu}+\hat{b}^{\dagger}_{\nu}\right)$
with $g_{\mathrm{opt},0}^{(N_{\mathrm{opt}})}=\sqrt{N_{\mathrm{opt}}}g_{\mathrm{opt},0}$ the collective optomechanical vacuum coupling rate and 
$\hat{a}^{\dagger}_{\mathrm{opt}},\hat{a}_{\mathrm{opt}}\ 
\left(\hat{b}^{\dagger}_{\nu},\hat{b}_{\nu}\right)$ the optical pump field bosonic ladder operators 
(the vibrational phononic operators at frequency $\nu$).
The number of molecules $N_{\mathrm{opt}}$ participating in the optomechanical interaction with the pump laser may be different from the number $N_{\mathrm{IR}}$ participating to IR absorption, as elaborated in Sec.~\ref{sec:molecule} below (cf. also Appendix~\ref{sec:DFT} Sec. 5,6). 

The optical antenna field can be split into an average coherent amplitude $\alpha$ and a fluctuating term 
so that $\hat{a}_{\mathrm{opt}}=\alpha+\delta\hat{a}_{\mathrm{opt}}$. Expanding to first order in $\alpha$ 
the optomechanical interaction we obtain the linearized interaction 
\begin{equation}
\hat{H}_{\mathrm{lin}}=-\hbar g_{\mathrm{opt},0}^{(N_{\mathrm{opt}})}\sqrt{\bar{n}_{\mathrm{opt}}}
\left(\delta\hat{a}^{\dagger}_{\mathrm{opt}}+\delta\hat{a}_{\mathrm{opt}}\right)
\left(\hat{b}_{\nu}+\hat{b}_{\nu}^{\dagger}\right),
\label{eq:lin}
\end{equation}
with $\bar{n}_{\mathrm{opt}}=|\alpha|^2$ 
the mean occupation of the optical antenna mode (see Appendix \ref{sec:OMframe}).

The spectral density of the output field on the optical port in [photons/(Hz$\cdot$s)] 
can be evaluated through the calculation of the two-time correlations 
of the optical output field operators \cite{wilson-rae_cavity-assisted_2008,scully_quantum_1997}: 
\begin{equation}
S^{\mathrm{out}}(\omega)\propto\frac{\kappa^{\mathrm{opt}}_{\mathrm{ex}}}{2\pi}
\int_{-\infty}^{\infty}
\mathrm{d}\tau\mathrm{e}^{i\omega \tau}\langle\delta\hat{a}_{\mathrm{opt}}^{\dagger}(\tau)\delta\hat{a}_{\mathrm{opt}}(0)\rangle.
\label{eq:noiseout}
\end{equation}
%where the prefactor in brackets takes into account the frequency dependence 
%of the radiative coupling rate of a dipolar emitter to the far-field. 

Following previous works in optomechanics \cite{schliesser2010cavity} and their extension to 
molecular optomechanics \cite{schmidt_quantum_2016,schmidt_linking_2017} 
we can write an analytical expression 
of the outgoing spectral density at the anti-Stokes 
sideband $S^{\mathrm{out}}(\omega_{\mathrm{aS}})\propto A^- \bar{n}_f/
\left(\Gamma_{\nu}^*+\Gamma_{\mathrm{opt}}\right)$ 
(at the Stokes sideband $S^{\mathrm{out}}(\omega_{\mathrm{S}})\propto A^+ \left(\bar{n}_f+1\right)/\left(\Gamma_{\nu}^*+\Gamma_{\mathrm{opt}}\right)$) 
with $\bar{n}_f$ the mean final phonon number of the vibrational mode.
Due to the IR and optomechanical interactions 
the intrinsic vibrational damping rate is modified 
$\Gamma_{\mathrm{tot}}=\Gamma_{\nu}^*+\Gamma_{\mathrm{opt}}$
with $\Gamma_{\nu}^*$ the IR antenna-assisted damping rate and 
$\Gamma_{\mathrm{opt}}=A^- - A^+$ the additional damping rate of electromagnetic origin 
characterized by the imbalance between the optical antenna-assisted 
transition rates to the ground $A^-$ and excited vibrational states $A^+$ (see Appendix \ref{sec:OMframe}).

The final phonon number in the vibrational mode, $\bar{n}_f$, is given by 
the expression \cite{wilson-rae_cavity-assisted_2008}
\begin{equation}
\bar{n}_f=\frac{\Gamma_{\nu}^*}{\Gamma_{\nu}^*+\Gamma_{\mathrm{opt}}}\bar{n}_{\mathrm{b}}
+\frac{A^+}{\Gamma_{\nu}^*+\Gamma_{\mathrm{opt}}}, 
\label{eq:popnu}
\end{equation} 
where $\bar{n}_{\mathrm{b}}= \bar{n}_{\mathrm{b}}^{\textsc{ir}} + \bar{n}_{\mathrm{th}} $ 
is the total phonon number in the absence of optical drive. 
It is the incoherent sum of the IR-induced vibrational excitation (Eq. \ref{eq:abs}) and the thermal noise, 
$\bar{n}_{\mathrm{th}} = 
1/\left(\exp\left[\hbar\omega_{\nu}/k_B T_{\text{bath}}\right]-1\right)$ for a bath temperature $T_{\text{bath}}$. 
We assume here that the pump laser does not lead to significant 
Ohmic heating of the system. 
It is however straightforward to model laser-induced heating 
by introducing a pump-power-dependent bath temperature $T_{\text{bath}}$. 

The resulting spectral density $S^{\mathrm{out}}_{\mathrm{opt}}$
in the absence of incoming IR radiation ($\bar{n}_{\mathrm{b}}^{\textsc{ir}}=0$)
should be integrated over the device's 
operational bandwidth ($\mathrm{BW}\equiv\Gamma_{\mathrm{tot}}$) 
to obtain its dark-count rate 
$\tilde{S}^{\mathrm{out}}_{\mathrm{opt}}=
\int_\mathrm{BW}S^{\mathrm{out}}_{\mathrm{opt}}\ \mathrm{d}\omega$. 
The dark-count rate arising from the thermal contribution to the first term in Eq. (\ref{eq:popnu})
can be reduced by cooling the bath, whereas the second term describes a minimal noise level 
resulting from phonon creation by spontaneous Stokes scattering of the pump laser, 
a process equivalent to quantum backaction in cavity optomechanics. 
Therefore an optimal power that maximizes the signal-to-noise ratio (SNR) exists, 
akin to the standard quantum limit (SQL) in position measurements.

From these expressions we are also able to describe the
conversion efficiency from an incoming rate of IR photons coupled to the antenna
into an outgoing rate of optical photons emitted by the antenna into free space, as 
$\tilde{S}^{\mathrm{out}}_{\textsc{ir}\mathrm{\rightarrow opt}}=\eta_{\mathrm{ext}}|\langle\hat{a}^\mathrm{in}_{\textsc{ir}}\rangle|^2$  
where $\eta_{\mathrm{ext}}=\eta_{\mathrm{opt}}\cdot\eta_{\mathrm{int}}\cdot\eta_{\textsc{ir}}$ 
is an external conversion efficiency \footnote{In this expression, we did not include the frequency dependence of the photonic density of state in free space, nor the factors related to the specific optical design used to couple light in and out of the structure.} 
with $\eta_{\mathrm{opt}}=\kappa^{\mathrm{opt}}_{\mathrm{ex}} / \kappa^{\mathrm{opt}}$.
%defined in the same way as $\eta_{\textsc{ir}}$ introduced previously. 

%These factors account for the coupling of free space radiation in and out of the nanostructure 
%\modif{where, for simplicity, we neglect the impact of the optics used on both these factors as well as the weak frequency dependence of $\kappa^{\mathrm{opt}}_{\mathrm{ex}}$.} 

%The radiative efficiency of the optical antenna mode into the far field is decomposed as 
%$\eta_{\mathrm{rad}}=\left(\omega/\omega_c\right)^3\eta_{\mathrm{opt}}$ 
%with $\eta_{\mathrm{opt}}=\kappa^{\mathrm{opt}}_{\mathrm{ex}} / \kappa^{\mathrm{opt}}$ 
%defined in the same way as $\eta_{\textsc{ir}}$ introduced previously. 
%These factors account for the coupling of free space radiation in and out of the nanostructure. 

The internal conversion efficiency $\eta_{\mathrm{int}}$ can in turn be divided into 
a power-dependent part $\eta_{\mathrm{OM}}(\bar{n}_\mathrm{opt})$ 
and a part describing the spatial overlap between the IR near field, 
the optical near field and the molecular ensemble, which we write $\eta_{\mathrm{overlap}}$. 
To approximate this last term we factorize it into two contributions: 
the spatial overlap between the IR and optical near fields ($\eta_{\mathrm{mode}}$) 
and the vectorial overlap between the near-field polarization (typically normal to the antenna surface) 
and the molecular orientation, which we name
$\eta_{\mathrm{pol}}$;
so that we can write 
\begin{equation}
\eta_{\mathrm{int}}\simeq\eta_{\mathrm{pol}}\cdot\eta_{\mathrm{mode}}
\cdot\eta_{\mathrm{OM}}(\bar{n}_\mathrm{opt}).
\label{eq:conv}
\end{equation}
The evaluation of $\eta_{\mathrm{overlap}}= \eta_{\mathrm{pol}} \eta_{\mathrm{mode}}$ is detailed in Appendix~\ref{sec:DFT} (Sec.~5,6).
The power and detuning dependence of the optomechanical 
efficiency term $\eta_{\mathrm{OM}}$ 
are depicted in Fig. \ref{SNR} (b) and its exact calculation is given in Appendix \ref{sec:OMframe}.

%%%%%%%%%%%%%%%%%%%%%%%%%%%%%%%%%%%%%%%%%%%%%%%%%%%%%%%%%%%%%%%%%%%%%%%%%%%%%%%%%%%%%%%%%%%%%%%%%%%%%%%%%%%%%%%%%%%%%%%%%%%%%%%%%%%%%%%%%%%%%%%%%%%%%%%%%%%%%%%%%%%%%%%%%%%%%%%%
\section{Molecular transducer}\label{sec:molecule}
The electric dipole moment $\vec{\mu}_{\nu}$ and polarizability $\alpha_{\nu}$ 
of a vibrational mode can be  extracted from experimental data of resonant light absorption and inelastic light scattering, respectively \cite{herzberg_molecular_1960,colthup_introduction_1990}. 
%In specific cases the symmetries of the vibrational mode lead to selection rules in its interaction with light \cite{wilson_molecular_1980}.
\begin{figure}[h!]
	\centering
	\hbox{\hspace{+6pt}
	\xincludegraphics[width=0.97\columnwidth,label= \textbf{\hspace{-9pt}(a)},fontsize=\large]{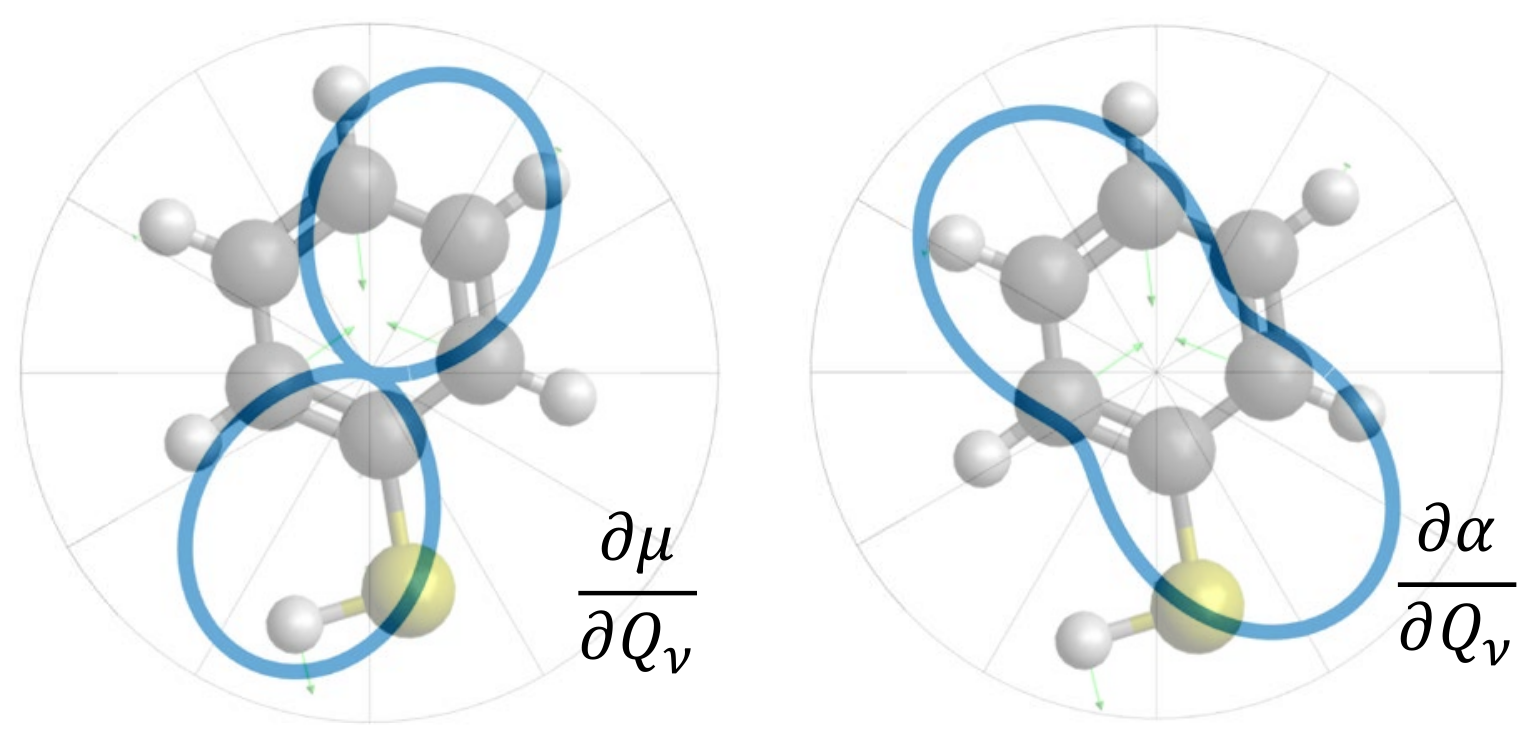}
	}
	\vspace{-10pt}
	\xincludegraphics[width=\columnwidth,label=\textbf{(b)},fontsize=\large]{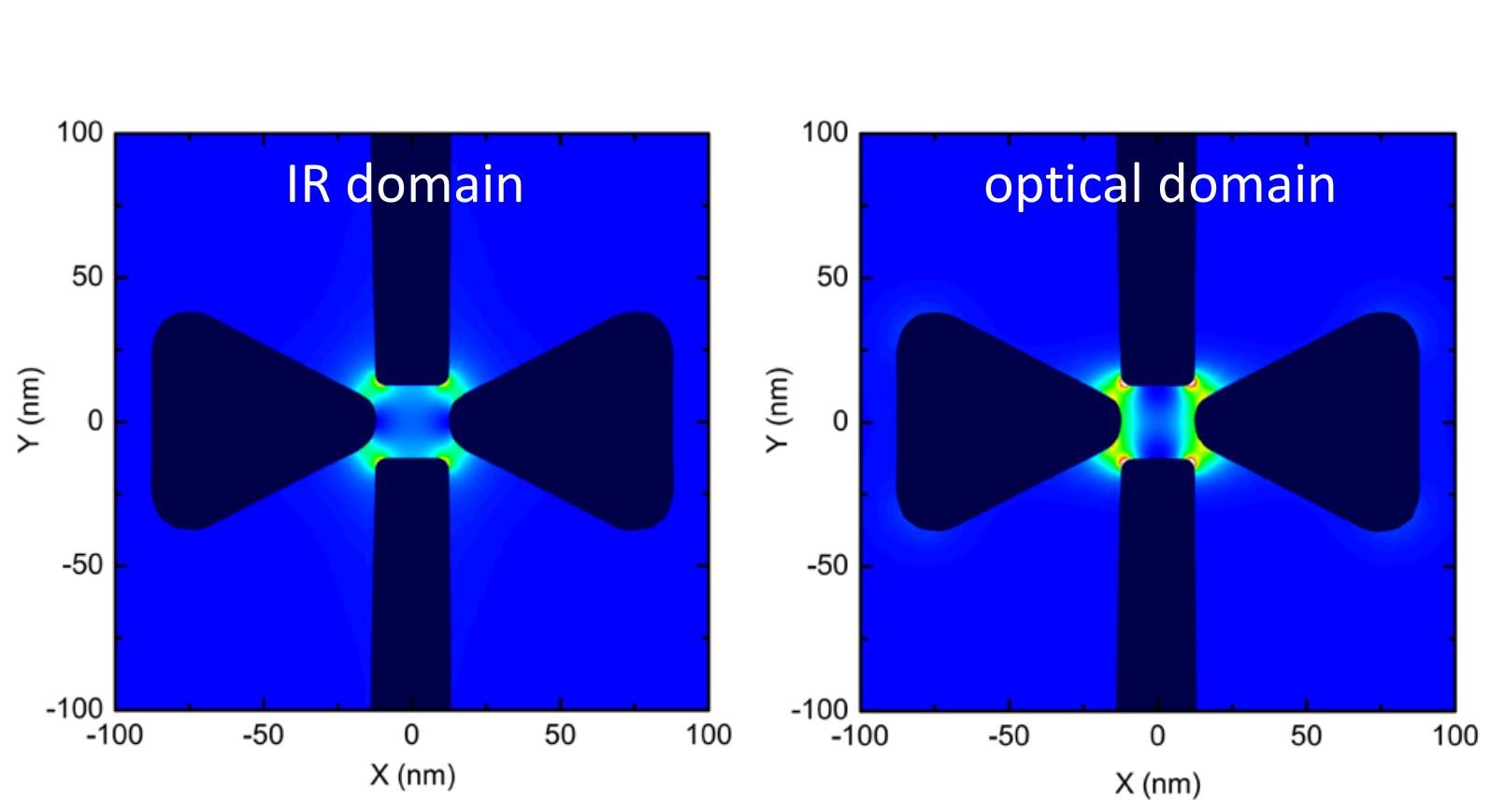}
	\vspace{-20pt}	
	\caption{
	\textbf{(a)} Polar plots of both 
	the electric moment derivative (left) 
	and the projection of the Raman tensor 
	on the main plane of the molecule (right) 
	for the vibrational mode $\tilde{\nu}=1002\ \mathrm{cm}^{-1}$ 
	of the thiophenol molecule (background image). 
	\textbf{(b)} Local density of states distribution inside the dual antenna for 
	an IR mode at 32 THz (left) and a NIR mode at 374 THz (right).
	}\label{THzAntenna}
	\vspace{-10pt}
\end{figure}
For vibrational modes lacking centrosymmetry the derivatives of 
both quantities with respect to the displacement coordinate can be nonvanishing \cite{wilson_molecular_1980}. 
We show such a situation in Fig.~\ref{THzAntenna} where we plot the projections 
of the derivatives of the electric moment and of the
polarizability with respect to the molecular coordinate $Q_{\nu}$ of the $1002\ \mathrm{cm}^{-1}$ mode of thiophenol, 
which we choose as an example in our calculations. 
We note that the projection of the tensor $\left(\frac{\partial\alpha_{\nu}}{\partial Q_{\nu}}\right)$ 
onto an axis perpendicular to the principal axis of the electronic moment derivative 
$\left(\frac{\partial\vec{\mu}_{\nu}}{\partial Q_{\nu}}\right)$ can be nonvanishing. 
Several polarization directions for in- and outcoupling of resonant and up-converted fields are thus conceivable.

The calculations leading to the parametric optomechanical vacuum coupling rate $g_{\mathrm{opt,0}}$ 
between the antenna field and the vibrational mode has been previously described \cite{roelli_molecular_2016} 
and its value is given by $g_{\mathrm{opt,0}}=\omega_c
\left(\vec{e}_{\mathrm{opt}}\cdot\frac{\partial\alpha_{\nu}}{\partial Q_{\nu}}\cdot\vec{e}_{\mathrm{opt}}\right)
\left(\frac{1}{V_{\mathrm{opt}}\varepsilon_0}\right)\sqrt{\frac{\hbar}{2\omega_{\nu}}}$ 
with $\alpha_{\nu}$ the polarizability tensor, $Q_{\nu}$ the reduced displacement coordinate 
of the vibrational mode labeled by $\nu$, $V_{\mathrm{opt}}$ the optical mode volume 
and $\vec{e}_{\mathrm{opt}}$ the unit polarization vector of the optical antenna mode.

The coupling rate $g_{\textsc{ir},0}$ associated with 
a vibrational mode $\nu$ is linked to an effective transition dipole  
$\vec{d}_{\nu}$ that can be numerically computed using e.g. density functional theory (DFT; cf. Appendix \ref{sec:DFT}).
Its value is
$g_{\textsc{ir},0}=\frac{1}{\hbar}\ \vec{d}_{\nu}\cdot\vec{\mathcal{E}}_0$
where the electric field per photon is given by 
$\vec{\mathcal{E}}_0=\sqrt{\frac{\hbar\omega_{\nu}}{2\varepsilon_0 V_{\textsc{ir}}}}\vec{e}_{\textsc{ir}}$ 
with $V_{\textsc{ir}}$ the mode volume and $\vec{e}_{\textsc{ir}}$ the unit polarization 
vector of the IR mode \cite{scully_quantum_1997,haroche_exploring_2006}.

Since $g_{\textsc{ir},0}^{(N_\textsc{ir})}$ scales with $\sqrt{N_{\textsc{ir}}/V_{\textsc{ir}}}$, 
it can be independent of the mode volume as long as this volume is filled with molecules. 
On the contrary the interaction of the vibration with the VIS-NIR optical field 
$g_{\mathrm{opt},0}^{(N_{\mathrm{opt}})}$ scales as 
$\sqrt{N_{\mathrm{opt}}}/V_{\mathrm{opt}}$ 
advocating for a device confining strongly this field, thereby reducing the  optical power required to reach an efficient conversion process.
  
%%%%%%%%%%%%%%%%%%%%%%%%%%%%%%%%%%%%%%%%%%%%%%%%%%%%%%%%%%%%%%%%%%%%%%%%%%%%%%%%%%%%%%%%%%%%%%%%%%%%%%%%%%%%%%%%%%%%%%%%%%%%%%%%%%%%%%%%%%%%%%%%%%%%%%%%%%%%%%%%%%%%%%%%%%%%%%%%  
\section{Dual plasmonic antenna}

Nanoantennas have proven 
to be instrumental in enhancing the interaction of molecules 
with off-resonant VIS-NIR optical fields (e.g. for surface-enhanced Raman scattering, SERS) 
\cite{zhu_quantum_2014,benz_single-molecule_2016}
and resonant IR fields (e.g. for surface-enhanced infrared absorption, SEIRA) 
\cite{brown_fan-shaped_2015,neubrech_surface-enhanced_2017}. 
We now present the design of a new dual-resonant antenna (see Figs.~\ref{FreqPic} and \ref{THzAntenna})
and compute the interaction of the local fields at the VIS-NIR and IR resonances with 
a specific vibrational mode of an ensemble of molecules covering the nanostructure. 
We assume that molecules are attached with their 
main axis perpendicular to the metallic surfaces, and extract from DFT calculations 
the relevant components of the derivatives of the electronic moment and polarizability with respect to the normal mode coordinate.
We note that calculations for specific self-assembled monolayer 
orientations \cite{schreiber_structure_2000,love_self-assembled_2005} 
or randomly oriented molecules can be achieved from the full knowledge 
of $\left(\frac{\partial\alpha_{\nu}}{\partial Q_{\nu}}\right)$ and 
$\left(\frac{\partial\vec{\mu}_{\nu}}{\partial Q_{\nu}}\right)$.

In our design the incoming field (to be up-converted) and the pump laser field 
are each resonant with a different component of the antennas arranged in a crossed configuration (cf. Appendix \ref{sec:antenna} for additional information on the design and values of its parameters).
At their intersection, the near-field polarizations of the two fields are collinear 
($\vec{e}_{\textsc{ir}}\simeq\vec{e}_{\mathrm{opt}}$),  
and we obtain $\eta_{\mathrm{pol}}=33$ \% for the vibrational 
mode illustrated in Fig.~\ref{THzAntenna}, when accounting for the specificities of the corresponding Raman tensor and IR transition dipole. 
Electromagnetic simulations demonstrate that the two fields, 
despite differing in frequency by more than one order of magnitude in that particular example, 
are confined within a very similar volume inside the nanogaps separating 
the two structures. This results in a spatial overlap of 
the two main electromagnetic field components 
within the dual antenna reaching $\eta_{\mathrm{mode}}=44$ \%.

From our numerical calculations we find that the antenna-assisted IR coupling rate 
for the vibrational mode at wavenumber ${\nu}=1002\ \mathrm{cm}^{-1}$
reaches $g_{\textsc{ir},0}^{(N_\textsc{ir})}/(2\pi)\sim 186$ GHz 
as $V_{\textsc{ir}}$ is decreased 
by several orders of magnitude below its diffraction limit 
(the calculation of $V_{\textsc{ir}}$ and $\vec{d}_{\nu}$ are detailed in Appendix~\ref{sec:abs}). 
As the cavity damping rate remains larger than the collective vacuum IR coupling rate 
($2g_{\textsc{ir},0}^{(N_\textsc{ir})}<\kappa^{\textsc{ir}}/2$ --- Purcell regime) 
the antenna-enhanced damping rate for this vibrational mode  
can be approximated by the expression 
\cite{haroche_exploring_2006} : 
$\Gamma_{\nu}^*\simeq\Gamma_{\nu}+\kappa^{\textsc{ir}}/2\left(1-
\sqrt{1-(2g_{\textsc{ir},0}^{(N_\textsc{ir})})^2/(\kappa^{\textsc{ir}}/2)^2}\right)$. 

We compute in Appendix \ref{sec:DFT} the coupling rate 
of another vibrational mode that has a larger IR dipole moment: 
Under optimal molecular orientation and filling conditions,
that mode is at the onset of the collective strong coupling regime 
with the IR antenna mode \cite{shalabney_coherent_2015}. 
While future work is needed to properly describe this regime 
in the context of wavelength conversion, 
we note that our design offers new perspectives to realize a source of IR photons. 
Indeed, in the strong coupling regime and under optomechanical parametric amplification 
\cite{roelli_molecular_2016}, the optically pumped population is shared 
between the collective vibrational mode and the corresponding IR antenna mode, 
resulting in the fast emission of IR radiation. 

%%%%%%%%%%%%%%%%%%%%%%%%%%%%%%%%%%%%%%%%%%%%%%%%%%%%%%%%%%%%%%%%%%%%%%%%%%%%%%%%%%%%%%%%%%%%%%%%%%%%%%%%%%%%%%%%%%%%%%%%%%%%%%%%%%%%%%%%%%%%%%%%%%%%%%%%%%%%%%%%%%%%%%%%%%%%%%%%
\section{Optical noise contributions}

\begin{figure}[h!]
	%\noindent
	\begin{minipage}{1.\columnwidth}
	\hspace{-12pt}
	\xincludegraphics[width=1.\columnwidth,label= \vbox{\vspace{0pt}\hbox{\hspace{-1pt}\textbf{(a)}}},
	fontsize=\large]{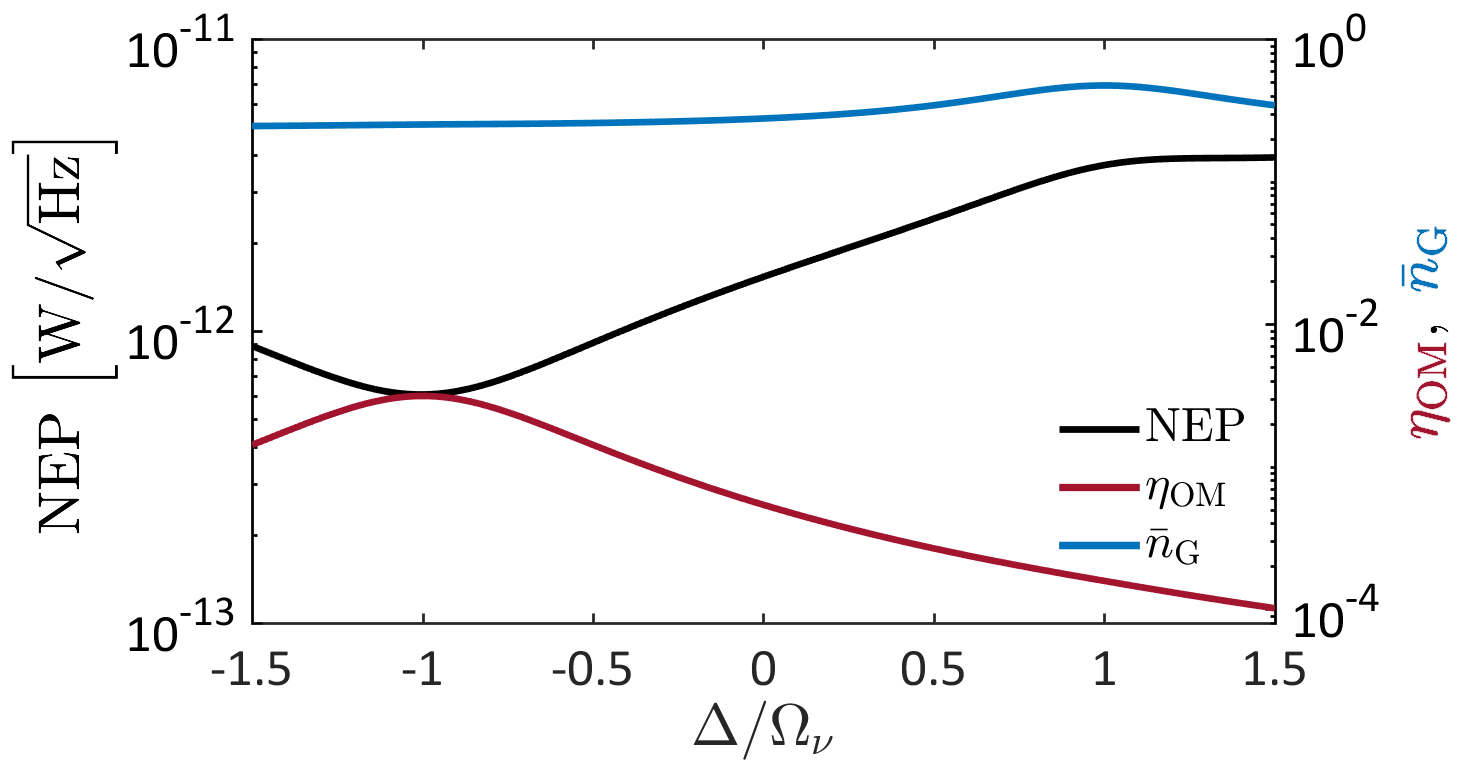}	
	\end{minipage}
	%\noindent
	\xincludegraphics[width=1.01\columnwidth,label=\vbox{\vspace{0pt}\hbox{\hspace{-5pt}\textbf{(b)}}},
	fontsize=\large]{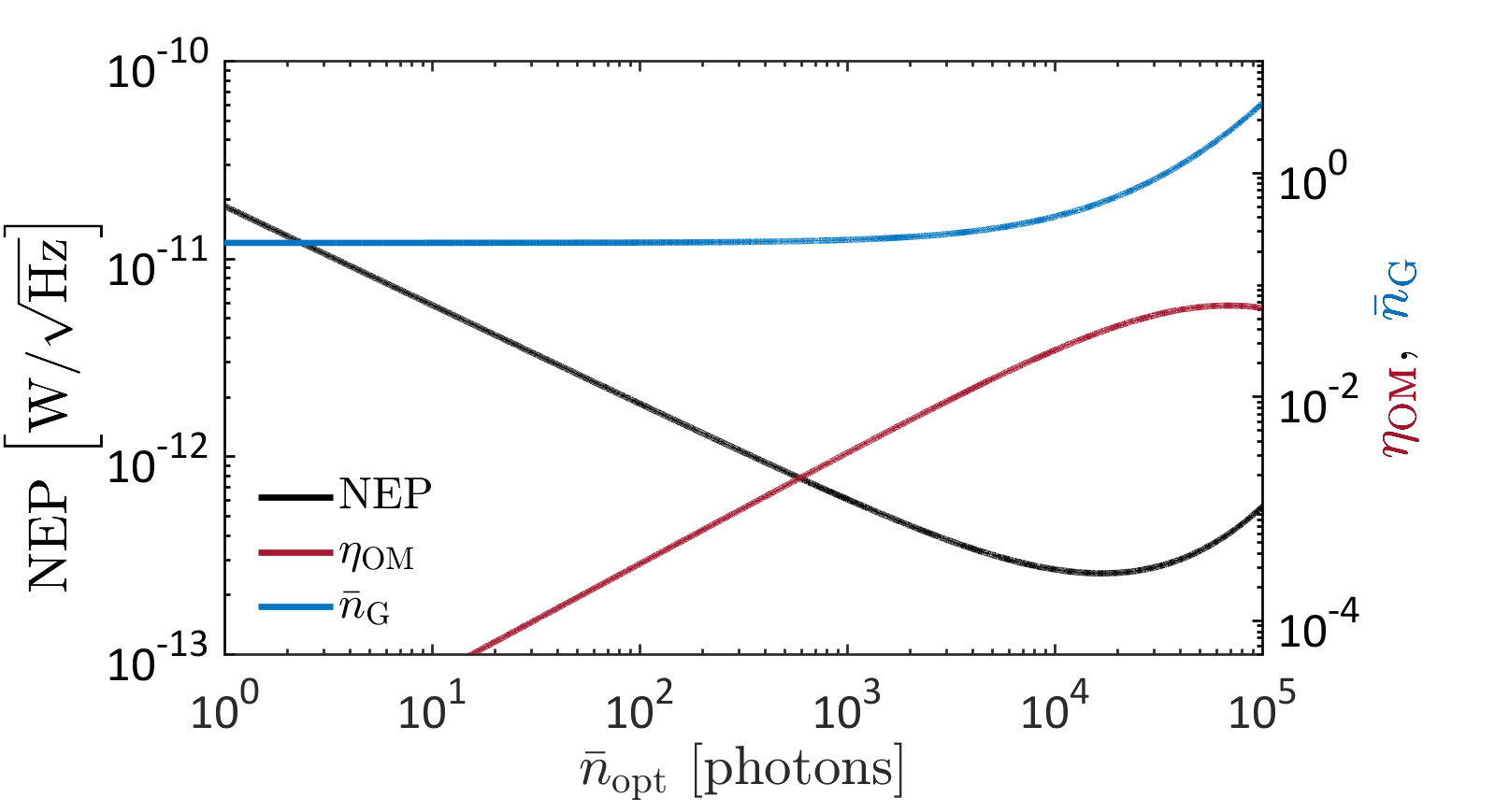}	
		
	\xincludegraphics[width=0.95\columnwidth,label=\hspace{-14pt} \textbf{(c)},
	fontsize=\large]{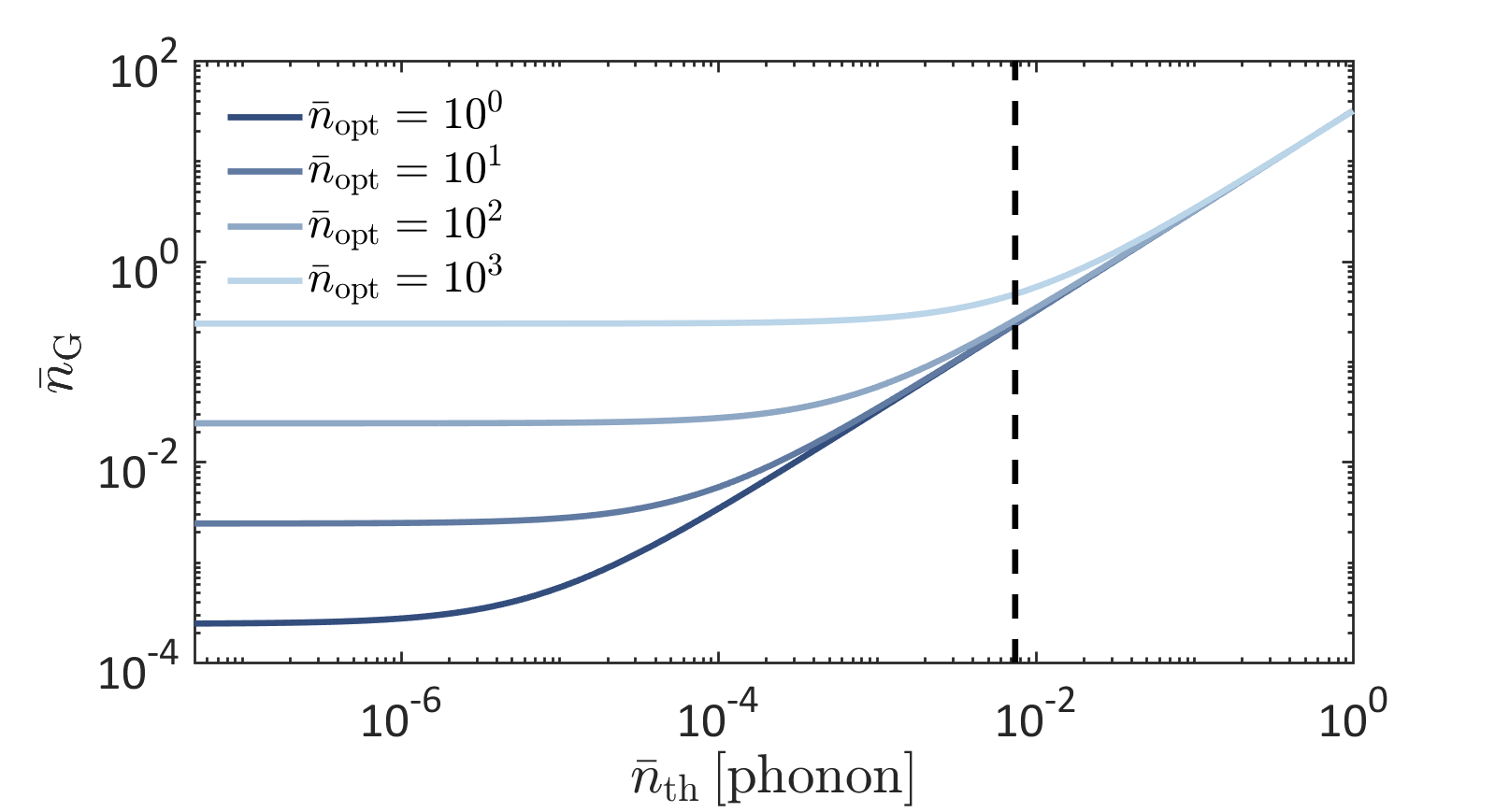}
	\vspace{-15pt}
	
	\caption{
	\textbf{(a-b)} 
	Left axis: Noise-equivalent power (NEP) (black solid line); 
	right axis: Power-dependent part of the internal conversion efficiency 
	$ \eta_{\mathrm{OM}}(\bar{n}_\mathrm{opt})$ (red solid line)
	and noise equivalent photon number per gate $\bar{n}_{\mathrm{G}}$ 
	(blue solid line) plotted on a logarithmic vertical scale
	\textbf{(a)} as a function of the detuning of the optical pump laser with respect to the plasmonic resonance  
	for a fixed intracavity optical photon number $\bar{n}_\mathrm{opt}=1000$;
	and  \textbf{(b)} plotted as a function of intracavity optical photon number $\bar{n}_\mathrm{opt}$ 
	for a fixed optimal red detuning $\Delta = - \Omega_\nu$. 
	We use the parameters for the dual antenna and molecular system described in the text. 
	All their numerical values can be found in the Appendixes.
	\textbf{(c)} $\bar{n}_{\mathrm{G}}$ 
	as a function of the thermal occupancy of the vibrational mode 
	for several intracavity optical photon numbers. The dashed vertical line denotes the thermal occupancy 
	of the mode considered in the text $\omega_{\nu}/(2\pi)=30$~THz at room temperature. 
}
	\label{SNR}
	\vspace{-10pt}
\end{figure}

A useful figure of merit to compare the performance of detectors independently of 
their respective operational bandwidth is the noise-equivalent power, $\mathrm{NEP}=P^{\mathrm{min, in}}_{\textsc{ir}}/\sqrt{\mathrm{BW}}\quad [\mathrm{W}\cdot\mathrm{Hz}^{-1/2}]$ 
where $P^{\mathrm{min, in}}_{\textsc{ir}}$ is the incoming power  at which the detector reaches 
a unity signal-to-noise ratio (SNR). This definition can be translated to a our converter by defining the   SNR as 
$\mathrm{SNR}(\omega)=\tilde{S}^{\mathrm{out}}_{\textsc{ir}\mathrm{\rightarrow opt}}
(\omega)/\tilde{S}^{\mathrm{out}}_{\mathrm{opt}}(\omega)$.
If we assume that the dark noise of a detector in the VIS-NIR range is negligible compared to the up-conversion noise, the NEP of the converter is by extension that of a detector built upon it.
%Assuming noise-less detection in the visible after the upconversion scheme suggested here, the NEP of our device can be evaluated. 
We show the computed NEP in Fig.~\ref{SNR} as a function of 
the optical pump power and laser detuning from the cavity resonance.
Remarkably, the NEP reaches values that improve on the state of the art for uncooled commercial devices (Table~\ref{tab:NEP}) 
and compares favorably with more recently demonstrated room-temperature platforms
\cite{long_room_2017,chen_widely_2017} operating at the higher end of the frequency range 
achievable with our molecular up-converter.

\begin{table}[h!]
\setlength\extrarowheight{8pt}
\begin{tabular}{l|c}
  \hline
  Detector type & NEP $\mathcal{O}\big(...\big)\ \left[\mathrm{W}\cdot\mathrm{Hz}^{-1/2}\right]$  
  \\[5pt] \hline
  Golay cell & $\mathcal{O}\Big(10^{-9}\Big)$ \tabularnewline
  Pyroelectric & $\mathcal{O}\Big(10^{-10}\Big)$ \tabularnewline
  MCT photodiode & $\mathcal{O}\Big(10^{-10}\Big)$ \tabularnewline
  Microbolometer & $\mathcal{O}\Big(10^{-11}\Big)$ \tabularnewline
  Molecular device & $\mathcal{O}\left(10^{-(11..12)}\right)$
\end{tabular}
\caption{Noise-equivalent power of commercially available uncooled devices
in the 5--50 THz region \cite{sizov_terahertz_2018,rogalski_two-dimensional_2019} 
and comparison with the device presented in this manuscript (molecular device).}
\label{tab:NEP}
\end{table}

When operating with a pump laser red-detuned from the optical resonance 
$\left(\Delta=\omega_p-\omega_c=-\omega_{\nu}\right)$ 
in the resolved sideband regime $\kappa^{\mathrm{opt}}/2<\omega_{\nu}$
we can simplify the interaction of Eq. \ref{eq:lin} and obtain 
\begin{equation}
\hat{H}_{\mathrm{eff}}=-\hbar g_{\mathrm{opt},0}^{(N_{\mathrm{opt}})}\sqrt{\bar{n}_{\mathrm{opt}}}
\left(\delta\hat{a}^{\dagger}_{\mathrm{opt}}\hat{b}_{\nu}+\mathrm{h.\ c.}
\right),
\label{eq:Heff}
\end{equation}
This regime provides maximal efficiency and optimal NEP, as seen in Fig.~\ref{SNR}(a). 
We note that for low-frequency vibrational modes the condition 
$\kappa^{\mathrm{opt}}/2<\omega_{\nu}$ could be achieved with the help of 
hybrid cavities that feature narrower linewidths 
\cite{doeleman_antennacavity_2016,gurlek_manipulation_2018}.

Keeping this optimal detuning, we investigate in Fig.~\ref{SNR}(b) 
how the NEP depends on optical pump power. 
As the intracavity pump photon number is increased, the efficiency initially grows linearly, 
while the noise remains constant, limited by the thermally generated anti-Stokes signal. 
This yields a square-root decrease of NEP with pump power. 
Interestingly, at high intracavity photon number the contribution of optomechanical 
quantum backaction to the dark-count rate surpasses the thermal contribution, 
and the NEP degrades with increasing power. This behavior is reminiscent of 
the standard quantum limit for displacement detection in optomechanical cavities.

Another unique feature of the up-conversion scheme is its compatibility with single-photon detectors already available in the VIS-NIR range. 
To assess more precisely the feasibility of operating our device in single-photon counting mode, 
we introduce the noise-equivalent photon rate, i.e. 
$\tilde{S}^{\mathrm{out}}_{\mathrm{opt}}/\eta_{\mathrm{ext}}\equiv
|\langle\hat{a}^\mathrm{in}_{\textsc{ir}}\rangle|^2/\mathrm{SNR}$. 
This quantity corresponds to the incoming IR photon rate 
at the input of the device that would generate an output rate of 
up-converted photons equal to the dark-count rate.

In practice, noisy single-photon detectors are best operated in gated mode. 
In our approach, this mode is easily realized using a pulsed optical pump laser, 
with a pulse duration $\Delta t$ of a few picoseconds that ideally matches the molecular vibrational linewidth; 
$\Delta t \simeq \left(\Gamma_{\nu}^{*}\right)^{-1}$. 
This mode of operation provides not only better noise rejection and higher intracavity photon numbers 
(therefore better efficiency), but also ultrafast timing resolution 
that is not otherwise achievable  due to the intrinsic timing jitter 
of VIS-NIR single-photon counters (typically several tens of picoseconds) \cite{hadfield_single-photon_2009}.  
We therefore define the noise-equivalent photon number per gate  
$\bar{n}_{\mathrm{G}}=\tilde{S}^{\mathrm{out}}_{\mathrm{opt}}
/\left(\eta_{\mathrm{ext}}\Gamma_{\nu}^{*}\right)$ 
which translates the noise-equivalent photon rate into an 
average incoming IR noise photons per time gate. 

%%%%%%%%%%%%%%%%%%%%%%%%%%%%%%%%%%%%%%%%%%%%%%%%%%%%%%%%%%%%%%%%%%%%%%%%%%%%%%%%%%%%%%%%%%%%%%%%%%%%%%%%%%%%%%%%%%%%%%%%%%%%%%%%%%%%%%%%%%%%%%%%%%%%%%%%%%%%%%%%%%%%%%%%%%%%%%%%
\section{Converter Arrays}
To gain more insight on the limiting factors constraining single-photon operation, 
we plot $\bar{n}_{\mathrm{G}}$ as a function of the thermal occupancy 
of the vibration and of the intracavity pump photon number in Fig.~\ref{SNR}(c). 
This graph shows that moderate cooling of the device to 100$^\circ$C below ambient
(achieved with thermoelectric cooling systems) would bring the thermal occupancy 
of this vibrational mode down to $\bar{n}_\text{th}=5.6 \cdot 10^{-4}$ allowing to reach a noise as low as $\bar{n}_{\mathrm{G}} \simeq 2 \cdot 10^{-2}$, 
making single-photon counting with picosecond time resolution 
in the MIR and FIR domains a realistic prospect. 
For the lower-frequency range (5--20 THz) reducing the temperature of the molecules
in a cryogenic environment would be required to allow single-photon operation. 

A promising way to further reduce the gated dark-count level consists 
in designing an array of molecular converters, 
sufficiently distant from each other so as not to interact by near-field coupling. 
We assume that the array is illuminated by spatially coherent IR signal and optical pump beam, 
which is achievable when using a high $f$-number lens 
due to the sub-wavelength dimensions of the antennas. 
The key advantage of this scheme is that the anti-Stokes fields of thermal origin 
from different antennas would not exhibit any mutual phase coherence; 
they will add up incoherently in the far field.
On the contrary, the up-converted (sum-frequency) anti-Stokes fields would be 
phase coherent and interfere constructively in specific directions, 
in analogy with a phased emitter array \cite{dregely_3d_2011,busschaert_beam_2019}. 
Considering a simple linear array, as demonstrated in Appendix \ref{sec:array}, 
this effect would jointly decrease the thermal contribution to the dark-count rate 
and dilute the intracavity photon number per device, enabling 
single-photon operation with improved sensitivity. 

A configuration with multiple converters within the IR spot 
could alternatively be leveraged for on-chip IR multiplexing 
\cite{schwarz_monolithically_2014,lin_mid-infrared_2018,
tittl_imaging-based_2018,yesilkoy_ultrasensitive_2019} 
with distinct converters responding to distinct IR frequencies by the proper choice of molecular vibrations and antenna design, thereby bypassing the limited detection bandwidth of a single converter. 
This sub-wavelength platform  benefits from the coherent nature of the conversion process and opens the route to IR spectroscopy, IR hyperspectral imaging and recognition technologies.\\

\section{Conclusion}

In summary, we presented a new concept for frequency upconversion from the mid-infrared to the visible domain based on the interaction of both fields with molecular vibrations coupled to a dual-resonant nanoantenna. 
We considered an incoming long-wavelength infrared radiation that resonantly excites a vibrational mode, 
which is simultaneously coupled through its Raman polarizability to a coherent pump field at shorter wavelength 
(visible or near-infrared), resulting in up-conversion of the IR signal onto the anti-Stokes sideband of the pump.  
Thanks to the recently developed framework of molecular cavity optomechanics, 
we were able to treat the problem in a full quantum model, 
and thereby predict the internal quantum efficiency of our device, 
as well as its outgoing noise spectral density. 
We showed that the noise added in the conversion process has two main origins: thermal noise and backaction noise (including quantum and dynamical backaction), the latter increasing superlinearly with pump power and eventually becoming dominant.
We analyzed the dependence of the noise-equivalent power (NEP)on the intracavity pump photon number and pump-cavity detuning, 
and predicted that under the optimal condition of red-sideband excitation, 
the NEP can be as low as few $\mathrm{pW}\cdot\mathrm{Hz}^{-1/2}$, 
{improving on the state of the art for devices operating at ambient conditions}.

We stress that our numerical estimates are based on a realistic nanoantenna design 
and a common simple molecule (thiophenol). 
Although the intracavity photon numbers {required to reach optimal performance appear to be large, 
they can be achieved under pulsed excitation \cite{albrecht2017}. 
Moreover requirements on the intracavity power would be lowered by further reducing the gap size 
(down to 1--2~nm) and {by chemical engineering of the molecular converter toward higher Raman activity}.
Our study also shows that by moderately increasing {the resonant coupling rate between molecular vibration and IR antenna}, 
the system would enter the IR strong coupling regime, 
with the formation of vibrational polaritons \cite{shalabney_coherent_2015}. 
We leave the study of the conversion process in this new regime for future investigation. 

%Code to reproduce the data of Figs.~\ref{SNR} and \ref{RSBconversion} 
%in addition to the antenna and molecular parameters is available 
%on Zenodo \cite{roelli_zenodo_2020}. 

\section*{Acknowledgements}
The authors thank Wen Chen and the reviewers of the article for valuable comments.
C. G. acknowledges support from the Swiss National Science Foundation 
through Grant No. PP00P2-170684.
This work has received funding from the European Union's Horizon 2020 research and innovation programme with grant agreement No. 732894 (HOT) and No. 829067 (THOR).
P. R. acknowledges support from the Max Planck-EPFL Center for Molecular Nanoscience and Technology. 
D. M.-C. thanks Vahid Sandoghdar and acknowledges financial support from the Max Planck Society. \\

\appendix
\section{Optomechanical framework}\label{sec:OMframe}

A detailed description of the optomechanical framework 
can be found elsewhere 
\cite{schliesser2010cavity,roelli_molecular_2016,schmidt_quantum_2016,schmidt_linking_2017}. Here we just remind the readers of the few definitions 
and relationships used in the paper. \\

\noindent The average number of intracavity excitations in the VIS-NIR (label `opt') or IR antenna mode is related to the incoming photon flux $|\langle\hat{a}^{\mathrm{in}}_{\mathrm{opt/\textsc{ir}}}\rangle|$ and incoming power $P^{\mathrm{in}}_{\mathrm{opt/\textsc{ir}}}$ by
\begin{widetext}
\begin{equation}
\bar{n}_{\mathrm{opt/\textsc{ir}}}=
|\langle\hat{a}^{\mathrm{in}}_{\mathrm{opt/\textsc{ir}}}\rangle|^2
\frac{\kappa_{\mathrm{ex}}^{\mathrm{opt/\textsc{ir}}}}
{\Delta^2+\left(\kappa^{\mathrm{opt/\textsc{ir}}}/2\right)^2}
=\frac{P^{\mathrm{in}}_{\mathrm{opt/\textsc{ir}}}}
{\hbar\omega_{p\mathrm{/\textsc{ir}}}}
\frac{\kappa_{\mathrm{ex}}^{\mathrm{opt/\textsc{ir}}}}
{\Delta^2+\left(\kappa^{\mathrm{opt/\textsc{ir}}}/2\right)^2}.
\end{equation}
\end{widetext}

When considering the molecular vibrational levels and their parametric coupling to the optical field, 
the antenna-assisted transition rate to a lower excited level (anti-Stokes transition) is given by
\begin{equation}
A^-=\left(g_{\mathrm{opt,0}}^{(N_\mathrm{opt})}\right)^2 \bar{n}_{\mathrm{opt}}
\frac{\kappa^{\mathrm{opt}}}{\left(\omega_{\nu}-\Delta\right)^2
+\left(\kappa^{\mathrm{opt}}/2\right)^2}.
\end{equation}
The antenna-assisted transition rate to a higher excited vibrational level (Stokes transition) is given by
\begin{equation}
A^+=\left(g_{\mathrm{opt},0}^{(N_\mathrm{opt})}\right)^2 \bar{n}_{\mathrm{opt}}
\frac{\kappa^{\mathrm{opt}}}{\left(\omega_{\nu}+\Delta\right)^2
+\left(\kappa^{\mathrm{opt}}/2\right)^2}.
\end{equation}

\noindent The interested reader can find the complete derivation of the outgoing spectral density 
in Ref. \cite{wilson-rae_cavity-assisted_2008}. In this manuscript we are interested in the signal 
arising on the anti-Stokes sideband. Starting from Eq.~(\ref{eq:noiseout}) in the main text 
and following the same calculation steps we arrive at the final expression 
\begin{equation}
S^{\mathrm{out}}_{\mathrm{tot}}(\omega_{\mathrm{aS}})=
\frac{2}{\pi}
\frac{\eta_{\mathrm{opt}}A^-}{\Gamma_{\nu}^*+\Gamma_{\mathrm{opt}}}
\bar{n}_{f}
\end{equation}

\noindent For convenience we label the different components of the outgoing noise spectral density 
according to the origin of the vibrational population from which they result (cf. Eq.~(\ref{eq:popnu}) in the main text):
\begin{widetext}
\begin{equation}
S^{\mathrm{out}}_{\mathrm{tot}}(\omega_{\mathrm{aS}})\propto
\underbrace{\frac{A^-}{\Gamma_{\nu}^*+\Gamma_{\mathrm{opt}}}
\bar{n}_{\mathrm{th}}}_{S^{\mathrm{out}}_{\mathrm{th}}}
+\underbrace{\frac{A^-\Gamma_{\nu}^*}{\left(\Gamma_{\nu}^*+\Gamma_{\mathrm{opt}}\right)^2}
\eta_{\mathrm{overlap}}\bar{n}_b^{\textsc{ir}}}_{S^{\mathrm{out}}_{\textsc{ir}\mathrm{\rightarrow opt}}}
+\underbrace{\frac{A^-}{\left(\Gamma_{\nu}^*+\Gamma_{\mathrm{opt}}\right)^2}
\left[A^+ - \Gamma_{\mathrm{opt}}\bar{n}_{\mathrm{th}}\right]}
_{S^{\mathrm{out}}_{\mathrm{ba}}}.
\label{eq:NSD}
\end{equation} 
\end{widetext}
With this notation the total noise quanta in the outgoing optical field is
 $S^{\mathrm{out}}_{\mathrm{opt}}=S^{\mathrm{out}}_{\mathrm{th}}+S^{\mathrm{out}}_{\mathrm{ba}}$.
\\\\
We can then derive the expression for the conversion efficiency defined as 
$\tilde{S}^{\mathrm{out}}_{\textsc{ir}\mathrm{\rightarrow opt}}
=\eta_{\mathrm{ext}}|\langle\hat{a}^\mathrm{in}_{\textsc{ir}}\rangle|^2$ 
and obtain 
\begin{equation}
\eta_{\mathrm{ext}}= \eta_{\mathrm{opt}}\cdot
\underbrace{\eta_{\mathrm{overlap}}\cdot
\frac{A^-\Gamma_{\nu}^*}{\Gamma_{\nu}^*+\Gamma_{\mathrm{opt}}}
\frac{1}{\kappa^{\textsc{ir}}}
\left(\dfrac{2g_{\textsc{ir},0}^{(N_\textsc{ir})}}{\Gamma_{\nu}^*+\Gamma_{\mathrm{opt}}}\right)^2}
_{\eta_{\mathrm{int}}}
\cdot\eta_{\textsc{ir}}.
\label{eq:eta_ext}
\end{equation} 
This expression comprises the different factors constituting Eq.~(\ref{eq:conv}) of the main text.
 
\noindent For a pump field red-detuned from the optical antenna resonance 
($\Delta=\omega_p-\omega_c=-\omega_{\nu}$) 
this expression can be further developed to evidence the dependence 
of the internal conversion efficiency on IR and optical collective vacuum coupling rates:
\begin{equation}
\eta_{\mathrm{int}}= \eta_{\mathrm{overlap}}\cdot
\dfrac{\left(2g_{\mathrm{opt},0}^{(N_\mathrm{opt})}\right)^2\bar{n}_{\mathrm{opt}}}
{\kappa^{\mathrm{opt}}\left(\Gamma_{\nu}^*+\Gamma_{\mathrm{opt}}\right)}
\frac{\Gamma_{\nu}^*}
{\left(\Gamma_{\nu}^*+\Gamma_{\mathrm{opt}}\right)}
\dfrac{\left(2g_{\textsc{ir},0}^{(N_\textsc{ir})}\right)^2}
{\kappa^{\textsc{ir}}\left(\Gamma_{\nu}^*+\Gamma_{\mathrm{opt}}\right)}.
\end{equation}

\noindent Using this conversion efficiency, we have an alternate way to calculate the NEP directly from the dark-count rate and efficiency of the device as 
$\mathrm{NEP}=\frac{\hbar\omega_{\nu}}{\eta_{\mathrm{ext}}}\sqrt{\tilde{S}^{\mathrm{out}}_{\mathrm{opt}}}$ 
\cite{hadfield_single-photon_2009}. 
This method gives identical results as the one presented in the main text.

In Fig.~\ref{RSBconversion} we show the computed conversion efficiency and  NEP for the case $\Delta=-\omega_{\nu}$. 
In this red-detuned configuration our model assumptions remain valid 
for a large range of optical intracavity photon numbers. 
At high optical power we observe that the efficiency and the NEP reach an extremal value when the backaction contribution to the outgoing noise equals that of thermal vibrations.

In Fig.~\ref{RSBconversion}b,c we also show how these extremal points move when changing the IR absorption cross-section (in b) or Raman activity (in c) of the molecules. As can be expected, improving the Raman activity of the molecules only displaces the curves to the left without modifying the extremal values -- meaning that lower optical pump power is required to reach the same NEP and conversion efficiency. 
In contrast, improving the IR absorption can lead to a lower achievable NEP, in a certain range. For too large IR cross-sections, the vibration starts to enter the strong coupling regime with the IR antenna, and a different treatment would be needed to provide accurate predictions. 

\begin{figure}[h!]
	\noindent
		\xincludegraphics[width=\columnwidth,label= \hspace{-5pt}\textbf{(a)},fontsize=\large]{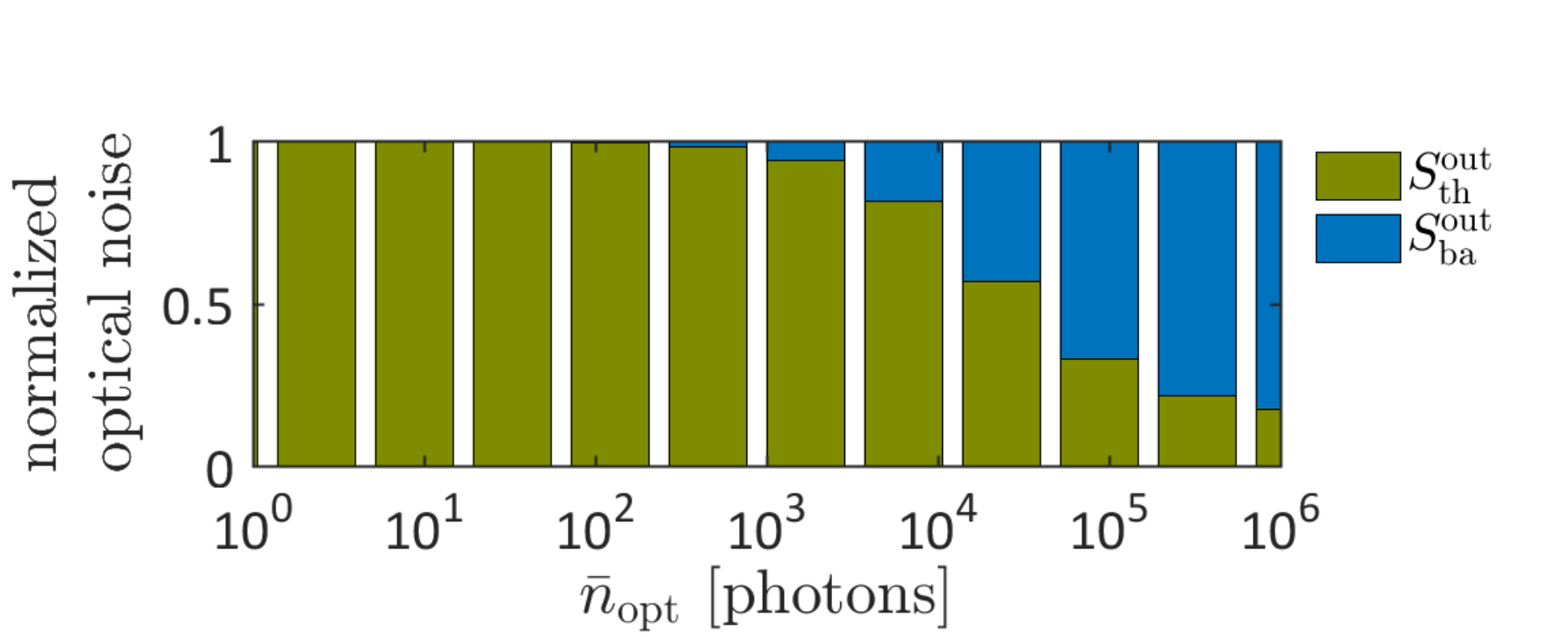}	
\hbox{\hspace{-8pt}
\xincludegraphics[width=0.93\columnwidth,label= \hspace{-10pt}\textbf{(b)},fontsize=\large]{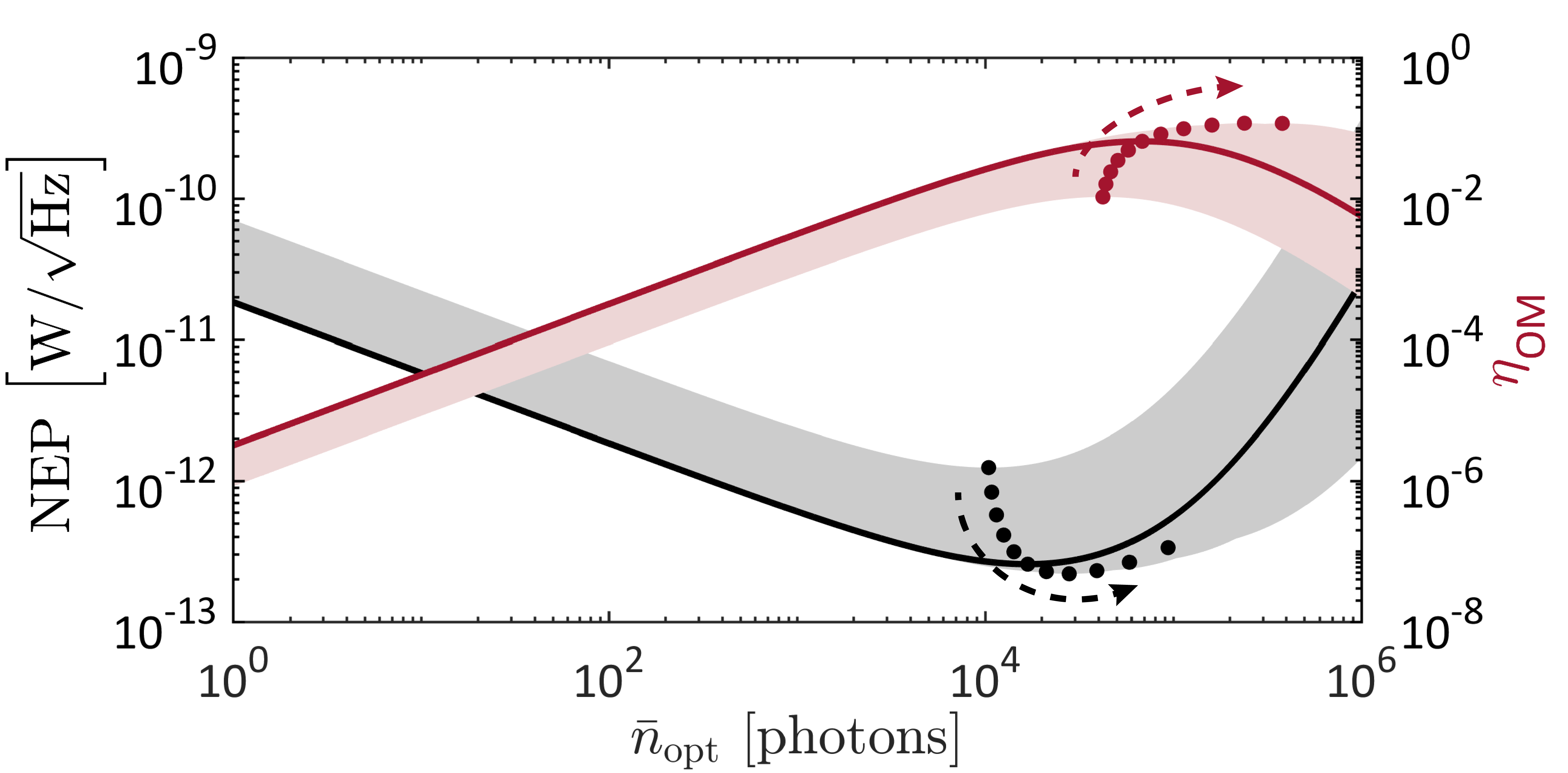}
} \hbox{\hspace{-6pt}
\xincludegraphics[width=0.93\columnwidth,label= \hspace{-10pt}\textbf{(c)},fontsize=\large]{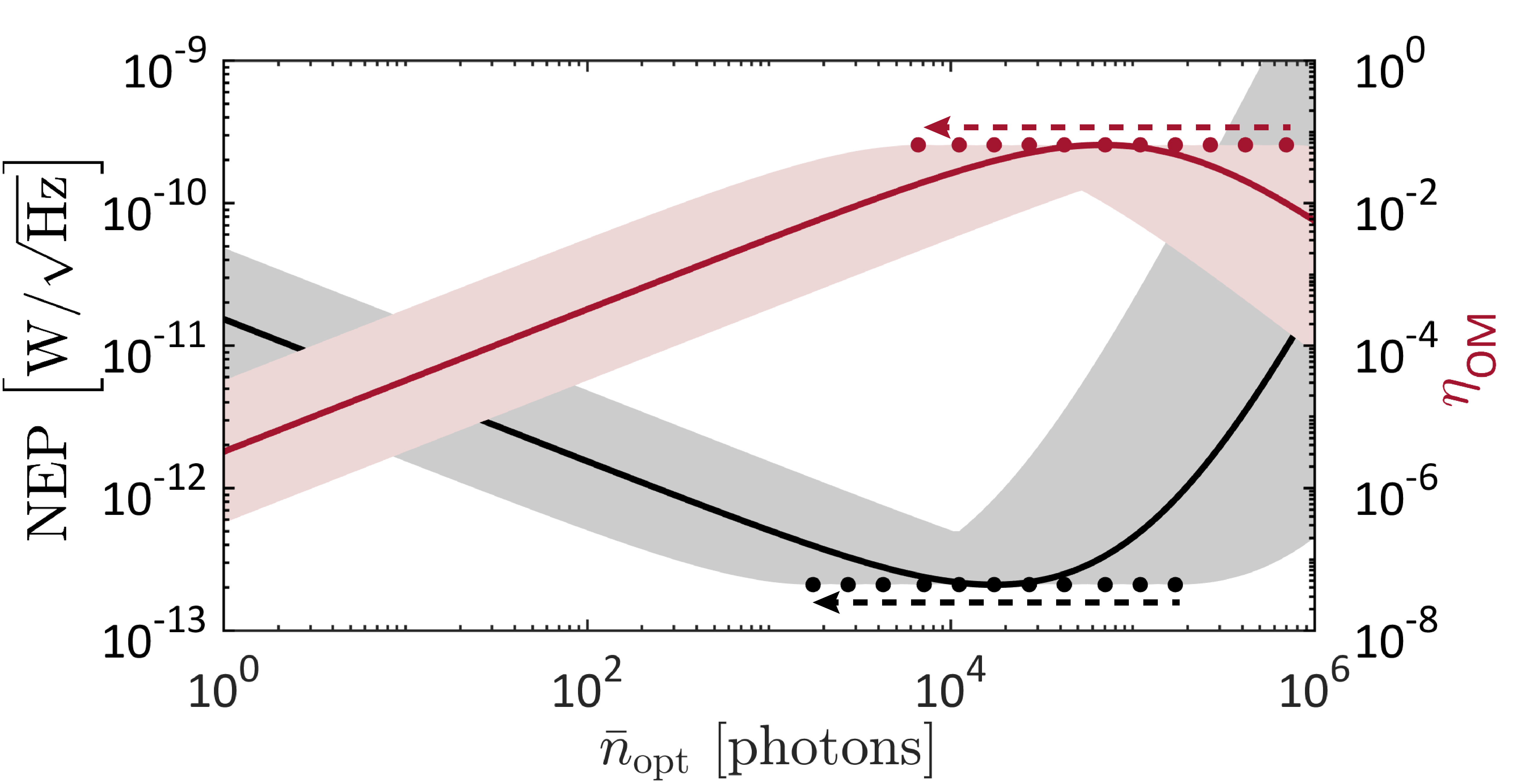}		
		}

	\vspace{-5pt}
	\caption{
	\textbf{(a)} Relative contributions to outgoing optical noise ($S^{\mathrm{out}}_{\mathrm{opt}}$)  
	on the anti-Stokes sideband as a function of intracavity photon number 
	for a pump tone red-detuned from the resonance of the optical antenna ($\Delta=-\omega_{\nu}$). 
	The contribution from the thermal population of the vibrational mode to the dark-count rate 
	is depicted in green and the backaction noise in blue.
	\textbf{(b-c)} 
	Noise-equivalent power ($\mathrm{NEP}$, left axis, black solid line) and 
	power-dependent part of the internal conversion efficiency 
	($\eta_{\mathrm{OM}}$, right axis, red solid line) 
	as a function of intracavity optical photon number for $\Delta=-\omega_{\nu}$.
	The parameters used to plot the lines are described in the Appendixes. 
	The dots indicate the extremal values in 	$\mathrm{NEP}$ and $\eta_{\mathrm{OM}}$ when varying 
	the absorption intensity $[0.1\mathrm{:}10]I_{\nu}^{\textsc{ir}}$ in \textbf{(b)} or 
	the Raman activity $[0.1\mathrm{:}10]R_{\nu}^{LL}$ in \textbf{(c)}. 
	The colored areas denote the $\mathrm{NEP}$ and $\eta_{\mathrm{OM}}$ achievable when sweeping 
	either parameter and the dashed arrows indicate the direction of evolution 
	of the extremum when increasing the parameter value. 
}
\vspace{-10pt}
	\label{RSBconversion}
\end{figure}

\section{Zero-temperature limit}
The limit of vanishing thermal occupancy of the vibrational mode is relevant for specific 
applications, and it demonstrates how the backaction noise acting onto the vibration sets a fundamental lower bound on the achievable NEP of the converter. 
When ($\bar{n}_{\mathrm{th}}\sim 0$) the 
outgoing noise spectral density of eq.~(\ref{eq:NSD}) 
can be simplified to 
\begin{equation} 
S^{\mathrm{out}}_{\mathrm{0K}}(\omega_{\mathrm{aS}})=
\frac{2}{\pi}\eta_{\mathrm{opt}}
\frac{A^- A^+}{\left(\Gamma_{\nu}^*+\Gamma_{\mathrm{opt}}\right)^2}.
\end{equation}
Taking into account eq.~(\ref{eq:eta_ext}) for the expression 
of the external conversion efficiency, the NEP 
and the $\bar{n}_{\mathrm{G}}$ at 0~K can be calculated. 

In the regime of weak optical pumping, 
$\Gamma_{\mathrm{opt}} < \Gamma_{\nu}^*$,
we obtain a linear scaling of
$\bar{n}_{\mathrm{G}}$ as a function of the intracavity 
photon number (appearing in the transition rates $A^-$ and $A^+$) 
while the NEP remains constant (cf. Fig.~\ref{Fig:0Kconversion}). 
The value of the NEP at this plateau, which corresponds to an intrinsic quantum limit due to measurement backaction, is given by the following expression 
\begin{equation}
\mathrm{NEP}_{\mathrm{0K}}=
\frac{\hbar\omega_{\nu} \kappa^{\textsc{ir}}}
{\eta_{\mathrm{overlap}}\sqrt{\eta_{\mathrm{opt}}}\eta_{\textsc{ir}}}
\sqrt{\frac{A^+}{A^-}}
\frac{\left(\Gamma_{\nu}^*+\Gamma_{\mathrm{opt}}\right)^{(5/2)}}
{\Gamma_{\nu}^*\left(2g_{\textsc{ir},0}^{(N_\textsc{ir})}\right)^2}.
\end{equation}

\begin{figure}[h!]
	\noindent
		\xincludegraphics[width=\columnwidth]{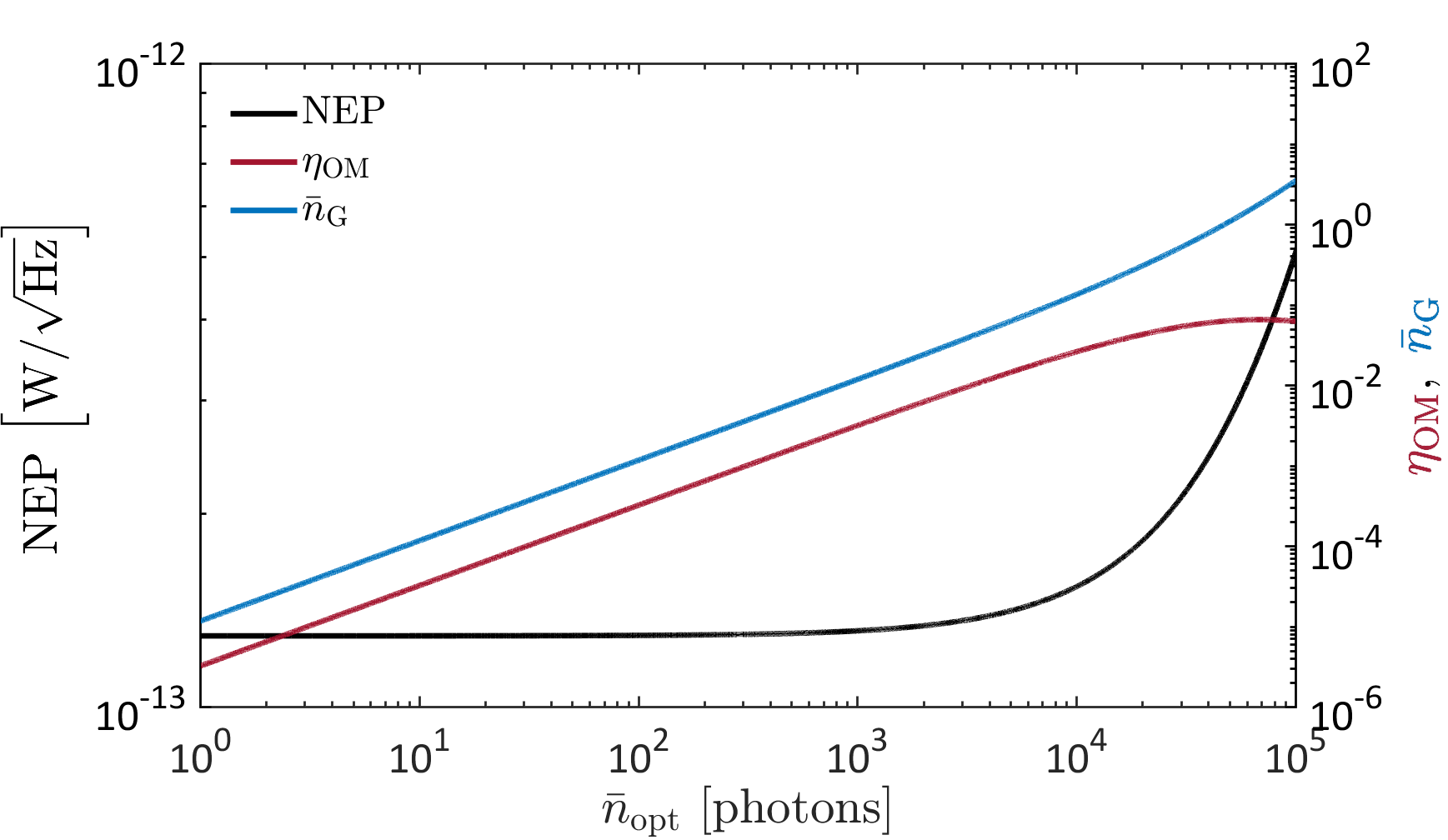}			
	\vspace{-5pt}
	\caption{
	Zero-temperature quantum limit on the NEP.
	Noise-equivalent power 
	($\mathrm{NEP}$, black solid line, left axis), 
	power-dependent part of the internal conversion efficiency 
	($\eta_{\mathrm{OM}}$, red solid line, right axis) 
	and noise-equivalent photon number per gate 
	($\bar{n}_{G}$, blue solid line, right axis) 
	as a function of the intracavity optical photon number 
	for $\Delta=-\omega_{\nu}$.
	The parameters used here are described in the Appendixes.
}
\vspace{-10pt}
	\label{Fig:0Kconversion}
\end{figure}

\section{Absorption of incoming IR radiation by molecular vibration}\label{sec:abs}

We describe in the following the coupling between a resonant field 
and a single molecular oscillator inside a cavity (i.e. antenna). 
We also derive the expression given in the main text for the number of excited phonons in the steady state. 
Our treatment is inspired by that of Ref. \cite{cohen-tannoudji_atom-photon_1992}.

The interaction between an external monochromatic field of frequency $\omega_{\textsc{ir}}$ and 
the molecular vibration inside a cavity is given in the dipole approximation by 
\begin{equation}
\hat{H}_{\mathrm{int}}=-\vec{\,d\,}\cdot\vec{\mathcal{E}}_{\textsc{ir}},
\end{equation}
where $\vec{\mathcal{E}}_{\textsc{ir}}=i\sqrt{\bar{n}_{\textsc{ir}}}
\left[e^{-i\omega_{\textsc{ir}}t}e^{-i\left(\phi+\phi_0\right)t}\vec{\mathcal{E}}_{0}-
e^{i\omega_{\textsc{ir}}t}e^{i\left(\phi+\phi_0\right)t}\vec{\mathcal{E}}^*_{0}\right]$  
with $\phi$ the phase offset between the field and the dipole, $\phi_0$ an adjustable 
phase parameter of the driving field. 
$\vec{\mathcal{E}}_0
=\sqrt{\frac{\hbar\omega_{\textsc{ir}}}{2\varepsilon_0 V_{\textsc{ir}}}}\vec{e}_{k}$ is the vacuum field, 
with $V_{\textsc{ir}}$ the mode volume and $\vec{e}_{k}$ the unit polarization vector of the IR mode.

This interaction can be written in terms of the bosonic ladder operators 
describing the IR mode inside the cavity $\hat{a}^{\dagger}_{\textsc{ir}},\hat{a}_{\textsc{ir}}$
and the vibrational phononic operators $\hat{b}^{\dagger}_{\nu},\hat{b}_{\nu}$ at frequency $\nu$. 
For a weak IR drive the vibrational Hilbert space of each molecule can be reduced to ground and first excited state 
$\{|0\rangle,|1\rangle\}$ 
and described like a two level system (TLS) with creation and annihilation operators 
$\hat{\sigma}^+_{\nu},\ \hat{\sigma}^-_{\nu}$ \cite{scully_quantum_1997}. 
We note that the validity of the TLS description 
for a collective vibrational mode of $N$ oscillators 
would break down only for a number of excitations of 
order $N$ \cite{haroche_exploring_2006}.

The dipolar transition is purely off-diagonal in this basis and described as 
$\vec{\,d\,}=d_{\nu}\left(
\hat{\sigma}^-_{\nu}\vec{e}_{\nu}+\hat{\sigma}^+_{\nu}\vec{e}_{\nu}^{\,*}\right)$.
The field inside the cavity is in turn described by 
$\vec{\mathcal{E}}_{\textsc{ir}}=-i\sqrt{\bar{n}_{\textsc{ir}}}\mathcal{E}_0
\left[ e^{-i\phi_0}\hat{a}_{\textsc{ir}}\vec{e}_{k}
- e^{i\phi_0}\hat{a}_{\textsc{ir}}^{\dagger}\vec{e}_{k}^{~*} \right]$ \cite{haroche_exploring_2006}.
In a frame rotating at the frequency of 
the IR driving field and keeping only the resonant processes 
($\hat{a}_{\textsc{ir}}^{\dagger}
\hat{\sigma}^-_{\nu},\hat{a}_{\textsc{ir}}\hat{\sigma}^+_{\nu}$)
 we obtain the interaction Hamiltonian :
\begin{equation}
\hat{H}_{\mathrm{int}}=-i\hbar g_{\textsc{ir}}\left(e^{-i\phi_0}\hat{a}_{\textsc{ir}}^{\dagger}
\hat{\sigma}^-_{\nu}+e^{i\phi_0}\hat{a}_{\textsc{ir}}\hat{\sigma}^+_{\nu}\right),
\end{equation}
with $g_{\textsc{ir}}
=\dfrac{d_{\nu}\cdot\mathcal{E}_0}{\hbar}\sqrt{\bar{n}_{\textsc{ir}}}\vec{e}_{\nu}^{\,*}\vec{e}_{k}e^{i\phi_0}
=g_{\textsc{ir},0}\sqrt{\bar{n}_{\textsc{ir}}}$.\\ 
Here we choose the additional phase term of the driving field 
in order for the coupling to be real positive, without loss of generality. 

We follow the dynamics of the TLS in this rotating frame. 
Introducing the rate $\Gamma_{\mathrm{tot}}$ which describes 
the total damping of the vibrational mode as described in the main text, we obtain 
\begin{subequations}
\begin{align}
\dree=& \quad i\frac{g_{\textsc{ir}}}{2} \left( \reo - \roe \right) - \Gamma_{\mathrm{tot}} \ree\\
\droo=& -i\frac{g_{\textsc{ir}}}{2} \left( \reo - \roe \right) + \Gamma_{\mathrm{tot}} \ree\\
\droe=& -i\delta\roe - i\frac{g_{\textsc{ir}}}{2} \left( \ree - \roo \right) - \frac{\Gamma_{\mathrm{tot}}}{2} \roe\\
\dreo=& \quad i\delta\reo + i\frac{g_{\textsc{ir}}}{2} \left( \ree - \roo \right) - \frac{\Gamma_{\mathrm{tot}}}{2} \reo,
\end{align}
\end{subequations}
with $\delta=\omega_{\textsc{ir}}-\omega_{\nu}$ the detuning between the IR drive 
and the vibrational resonance. \\

\noindent These equations are often described with the help of the Bloch vector components:
\begin{subequations}
\begin{align}
u=&\frac{1}{2}\ \left( \roe + \reo \right)\\
v=&\frac{1}{2i}\left( \roe - \reo \right)\\
w=&\frac{1}{2}\ \left( \ree - \roo \right).
\end{align}
\end{subequations}

The components $u,v$ of the Bloch vectors are related to the average dipole value 
\cite{cohen-tannoudji_atom-photon_1992}: 
$\langle\vec{\,d\,}\rangle=2\vec{d}_{\nu}
\left(u\cos\omega_{\textsc{ir}} t - v\sin\omega_{\textsc{ir}} t \right)$. 
We derive the master equations as a function of these components: 
\begin{subequations} 
\begin{align}
\dot{u}=& \quad \delta \, v - \frac{\Gamma_{\mathrm{tot}}}{2} u\\
\dot{v}=& -\delta \, u - g_{\textsc{ir}} w - \frac{\Gamma_{\mathrm{tot}}}{2} v\\
\dot{w}=& \quad g_{\textsc{ir}} v - \Gamma_{\mathrm{tot}} w - \frac{\Gamma_{\mathrm{tot}}}{2}.
\end{align}
\end{subequations}

\noindent The steady-state solutions of these equations are 
\begin{subequations}
\begin{align}
\bar{u}\quad &=\frac{g_{\textsc{ir}}}{2}
	\frac{\delta}{\delta^2+\left(\Gamma_{\mathrm{tot}}^2/4\right)+\left(g_{\textsc{ir}}^2/2\right)}\\
\bar{v}\quad &=\frac{g_{\textsc{ir}}}{2}
	\frac{\Gamma/2}{\delta^2+\left(\Gamma_{\mathrm{tot}}^2/4\right)+\left(g_{\textsc{ir}}^2/2\right)}\\
\bar{w}+\frac{1}{2}&=\frac{g_{\textsc{ir}}^2}{4}
	\frac{1}{\delta^2+\left(\Gamma_{\mathrm{tot}}^2/4\right)+\left(g_{\textsc{ir}}^2/2\right)}.
\end{align}
\end{subequations} \\

The average number of photons absorbed per unit time by the vibrational dipole is given by 
\begin{equation}
\frac{\mathrm{d}\bar{n}_{\mathrm{b}}^{\textsc{ir}}}{\mathrm{d}t}= 
\frac{\mathrm{d}\bar{W}^{\textsc{ir}}}{\mathrm{d}t}\frac{1}{\hbar \omega_{\textsc{ir}}}= 
\frac{\vec{\mathcal{E}}_n\cos\omega_{\textsc{ir}} t\cdot\langle\dot{\vec{\,d\,}}\rangle}
{\hbar \omega_{\textsc{ir}}}.
\end{equation}
If the detuning and coupling are much smaller than the vibrational damping rate (${\delta,g_{\textsc{ir}}} < \Gamma_{\mathrm{tot}}$), 
the average number of absorptions over an IR period can be written as  
\begin{equation}
\frac{\mathrm{d}\bar{n}_{\mathrm{b}}^{\textsc{ir}}}{\mathrm{d}t}= 
g_{\textsc{ir}}\bar{v} = \frac{g_{\textsc{ir}}^2}{2}
	\frac{\Gamma_{\mathrm{tot}}/2}{\delta^2+\left(\Gamma_{\mathrm{tot}}^2/4\right)+\left(g_{\textsc{ir}}^2/2\right)}
	\simeq\frac{g_{\textsc{ir}}^2}{\Gamma_{\mathrm{tot}}}.
\end{equation}
In the steady state the rate of photons absorbed by the vibrational mode 
equals the phonon damping rate so that the average number of excited phonons is 
\begin{widetext}
\begin{equation}
\bar{n}_{\mathrm{b}}^{\textsc{ir}}=\frac{\mathrm{d}\bar{n}_{\mathrm{b}}^{\textsc{ir}}}{\mathrm{d}t} 
\frac{1}{\Gamma_{\mathrm{tot}}} = \frac{g_{\textsc{ir}}^2}{\Gamma_{\mathrm{tot}}^2} 
= \frac{g_{\textsc{ir},0}^2}{\Gamma_{\mathrm{tot}}^2}\bar{n}_{\textsc{ir}}
= \frac{g_{\textsc{ir},0}^2}{\Gamma_{\mathrm{tot}}^2} 
\frac{\kappa^{\textsc{ir}}_{\mathrm{ex}}}{\delta^2+\left(\kappa^{\textsc{ir}}/2\right)^2}
\tilde{S}_{\textsc{ir}}^\mathrm{in}
\overset{\delta\ll\kappa^{\textsc{ir}}}{=}\frac{4g_{\textsc{ir},0}^2}{\Gamma_{\mathrm{tot}}^2}
\frac{\eta_{\textsc{ir}}}{\kappa^{\textsc{ir}}}
|\langle\hat{a}_{\textsc{ir}}^\mathrm{in}\rangle|^2.
\end{equation}
\end{widetext}
The condition $\delta\ll\kappa^{\textsc{ir}}$ is satisfied in realistic scenarios since the IR antenna decay rate $\kappa^{\textsc{ir}}$ is typically faster than the vibrational damping rate $\Gamma_{\mathrm{tot}}$. 
We note that the average number of excited phonons 
$\bar{n}_{\mathrm{b}}^{\textsc{ir}}$ 
can also be simply derived from the steady-state population of the upper TLS state 
$\bar{n}_{\mathrm{b}}^{\textsc{ir}}=\bar{w}+\frac{1}{2}$.

\section{Simulation of molecular parameters}\label{sec:DFT}
\label{IRabs}
The infrared absorption intensity of fundamental vibrational transitions 
$I_{\nu}^{\textsc{ir}}$ can be obtained by DFT calculations \cite{wilson_molecular_1980,feugmo_analyzing_2013} of the derivatives of the electric moment components $\mu_{\nu}^i$ 
with respect to the normal coordinates representing the vibrational mode of interest. 
They are usually expressed in [km$\cdot$mol$^{-1}$]. 
For a non-degenerate and harmonic vibrational mode the absorption intensity 
averaged over all orientations is given by 
\begin{equation}
\langle I_{\nu}^{\textsc{ir}}\rangle=\frac{N_A}{12\varepsilon_0c^2}
\sum_{i=1}^3 \left(\frac{\partial \mu_{\nu}^i}{\partial Q_{\nu}}\right)^2,
\end{equation}
with $N_A$ the Avogadro number.\\

\subsection{\textsc{Gaussian} calculations} 
The procedure is well described in the context of Raman calculations 
in the book of Le Ru \& Etchegoin \cite{ru_principles_2008}. 
The \textsc{Gaussian} software gives access to the derivatives of the electric dipole 
with respect to the \textit{i}th component of the displacement 
in Cartesian coordinates of the \textit{n}th atom. 
These derivatives can then be converted to derivatives with respect 
to the normal coordinates of a vibrational mode $\nu$ and are given in atomic units, 
i.e. the electric moment 
is given in bohr-electron (2.54 D / $8.48\cdot10^{-30}$ C$\cdot$m) 
and the displacement in bohr (0.529 \AA). 
The quantities $\frac{\partial \vec{\mu}_{\nu}}{\partial Q_{\nu}}$ 
can be converted to other systems of units:
\begin{widetext}
\begin{equation}
\left(\frac{\partial \vec{\mu}_{\nu}}{\partial Q_{\nu}}\right)^2 [\mathrm{D}^2\cdot\mathrm{\AA}^{-2}
\cdot\mathrm{amu}^{-1}]=\left[\frac{2.54}{0.53}\right]^2\cdot\left(\frac{\partial \vec{\mu}_{\nu}}{\partial Q_{\nu}}\right)^2 [\mathrm{atomic\ units}],
\end{equation}
\end{widetext}
and finally linked to the absorption intensity $I_{\nu}^{\textsc{ir}}$ 
of an incoming field of polarization $\vec{e}_{i}$ 
\begin{equation}
I_{\nu}^{\textsc{ir}}[\mathrm{km}\cdot\mathrm{mol}^{-1}]=126.8\cdot 
\left(\vec{e}_{i}\cdot\frac{\partial \vec{\mu}_{\nu}}{\partial Q_{\nu}}\right)^2 [\mathrm{D}^2\cdot\mathrm{\AA}^{-2}\cdot\mathrm{amu}^{-1}].
\end{equation}

\begin{table*}[t!]
\setlength\extrarowheight{8pt}
\begin{tabular}{l|c|c|c|c|c}
\hline
  Mode $[\mathrm{cm}^{-1}]$ & $I_{\nu}^{\textsc{ir}} 
  \left(\langle I_{\nu}^{\textsc{ir}} \rangle\right) 
  [\mathrm{km}\cdot\mathrm{mol}^{-1}]$ 
  & $R_{\nu}^{LL} 
  \left(\langle R_{\nu}^{LL}\rangle \right)
  [\mathrm{\AA}^4\cdot\mathrm{amu}^{-1}]$ 
  & $\eta_{\mathrm{pol}} 
  \left(\langle \eta_{\mathrm{pol}} \rangle\right) $
  & Coverage
  & $g_{\textsc{ir},0}^{(N_\textsc{ir})}/\kappa^{\textsc{ir}} $
 \\[5pt] \hline
 \multirow{2}{*}{mode 1002} & \multirow{2}{*}{0.52 (0.51)} & \multirow{2}{*}{2.40 (0.96)}
 & \multirow{2}{*}{0.33 (0.14)} & monolayer & 0.01\\\cline{5-6}
 & & & & volume & 0.06\\\cline{5-6}
 \hline
 \multirow{2}{*}{mode 1093} & \multirow{2}{*}{86.95 (28.52)} & \multirow{2}{*}{0.85 (0.31)} 
 & \multirow{2}{*}{0.97 (0.18)} & monolayer & 0.17\\\cline{5-6}
 & & & & volume & 0.75\\\cline{5-6}
\hline
\end{tabular}
\caption{Molecular parameters of interest for our conversion scheme 
for two vibrational modes of the thiophenol molecule. Calculations are obtained for a molecule 
oriented vertically with respect to both IR and VIS/NIR local fields 
(values averaged over all molecular orientations are given in parentheses for completeness). 
The resulting resonant coupling terms are calculated for 
two different coverages of the nanostructure by the molecules and given in units of $\kappa^{\textsc{ir}}$.}
\label{tab:DFTparam}
\end{table*}

\subsection{Cross-section} 
This expression can be linked to the absorption cross-section 
$I_{\nu}^{\textsc{ir}}=N_A\int\sigma_{\nu',\mathrm{abs}}\ \mathrm{d}\nu'$. 
If we assume a Lorentzian profile for the transition considered, 
the on-resonance value of the cross-section is $\int\sigma_{\nu',\mathrm{abs}}\ \mathrm{d}\nu'=
\frac{\pi}{2}\sigma_{\mathrm{abs}}\left(\nu'=\nu\right)\delta\nu=
\frac{\pi}{2}\sigma_{\mathrm{abs}}\left(\nu'=\nu\right)\frac{\Gamma_{\mathrm{tot}}}{2\pi c}$.

\noindent Thus, the absorption cross-section can be inferred from DFT calculations:
\begin{equation}
\sigma_{\nu,\mathrm{abs}}\ [\mathrm{cm}^2]= \frac{4c}{N_A\Gamma_{\mathrm{tot}}}I_{\nu}^{\textsc{ir}}\ 
[\mathrm{km}\cdot\mathrm{mol}^{-1}]\cdot 10^7.
\end{equation}

\subsection{Effective dipole moment} 
Accordingly we can also describe an effective dipole moment 
$d_{\nu}$ to characterize the vibrational transition and 
link it explicitly to the electronic moment derivatives found in DFT calculations:
%link it to the value of the absorption cross-section:
\begin{equation}
d_{\nu}=\sqrt{\frac{3\hbar\varepsilon_0 c\Gamma_{\mathrm{tot}}\sigma_{\nu,\mathrm{abs}}}{2\omega_{\nu}}}
=\sqrt{\frac{6\hbar\varepsilon_0 c^210^3}{N_A}}\sqrt{\frac{I_{\nu}^{\textsc{ir}}[\mathrm{km}\cdot\mathrm{mol}^{-1}]}{\omega_{\nu}}}
%=9.14\cdot10^{-25}\sqrt{\frac{I_{\nu}^{\textsc{ir}}}{\omega_{\nu}}}.
\end{equation} 

\subsection{Raman activity of an ensemble of molecules}
We refer the interested reader to Refs.  \cite{ru_principles_2008,roelli_molecular_2016} for 
detailed descriptions of the Raman activity, its connection 
with the optomechanical coupling rate and its calculation through DFT. 
For completeness we reproduce here a few expressions 
of the tensorial quantity $\frac{\partial \alpha_{\nu}}{\partial Q_{\nu}}$ 
averaged over randomly oriented molecules.  
To simplify the notation we introduce the Raman tensor 
$R_{\nu}=\frac{\partial \alpha_{\nu}}{\partial Q_{\nu}}$ and we refer to the scalar 
$\left(\vec{e}_{i}\cdot R_{\nu}\cdot\vec{e}_{j}\right)$ as $R_{\nu}^{ij}$. 
Taking $\vec{e}_{i} \perp \vec{e}_{j}$ and 
averaging over random orientations of the molecules one can obtain 
\begin{align}
\langle\left|R_{\nu}^{ii}\right|^2\rangle=
&\sqrt{\frac{45\bar{\alpha_{\nu}}^2+4\bar{\gamma_{\nu}}^2}{45}}\\
\langle\left|R_{\nu}^{ji}\right|^2\rangle=
&\sqrt{\frac{3\bar{\gamma_{\nu}}^2}{45}},
\end{align}
with $\bar{\alpha}_{\nu}^2=\frac{1}{9}\left[R_{\nu}^{xx}+R_{\nu}^{yy}+R_{\nu}^{zz}\right]^2$ 
and $\bar{\gamma}^2_{\nu}=
\frac{1}{2}\left[\left(R_{\nu}^{xx}-R_{\nu}^{yy}\right)^2+
\left(R_{\nu}^{yy}-R_{\nu}^{zz}\right)^2+
\left(R_{\nu}^{zz}-R_{\nu}^{xx}\right)^2\right]
+3\left[\left(R_{\nu}^{xy}\right)^2+\left(R_{\nu}^{xz}\right)^2+\left(R_{\nu}^{yz}\right)^2\right]$. 
These quantities do not depend on the two orthogonal orientations of the field 
chosen as polarization basis but only on the intrinsic properties of the molecule. 
In that situation Raman scattering can be described by a scalar named 
the magnitude of the Raman tensor 
$\mathcal{R}^2=\langle\left|R_{\nu}^{ii}\right|^2\rangle+
\langle\left|R_{\nu}^{ji}\right|^2\rangle
=\frac{45\bar{\alpha}_{\nu}^2+7\bar{\gamma}_{\nu}^2}{45}$ 
and can be derived directly from DFT calculations. 

\noindent We also introduce the depolarization ratio 
$\rho=\langle\left|R_{\nu}^{ji}\right|^2\rangle/
\langle\left|R_{\nu}^{ii}\right|^2\rangle$ 
that evaluates the importance of the cross-polarized component of
 the Raman-scattered field (with respect to the incoming field) and that is bounded by $0\leq\rho\leq3/4$. 
In the SERS scenario the outgoing field is solely polarized along the direction 
of the local cavity field $\vec{e}_{L}$. 
For randomly oriented molecules the magnitude of the Raman tensor is thus rescaled 
by a factor dependent on the depolarization ratio: 
\begin{equation}
\langle\left|R_{\nu}^{LL}\right|^2\rangle
=\mathcal{R}^2\frac{1}{1+\rho},
\end{equation} 

\subsection{Local overlap - $\eta_{\mathrm{pol}}$}
The factor $\eta_{\mathrm{pol}}$ describes the local overlap between the two fields 
involved in our conversion scheme, on the one hand, 
and the IR dipole and Raman tensor of the molecular vibration, on the other hand. 
It is defined in the following way: 
\begin{equation}
\eta_{\mathrm{pol}}=
\frac{\vec{e}_{\mathrm{L}}\cdot
\frac{\partial \vec{\mu}_{\nu}}{\partial Q_{\nu}}
}
{\|\frac{\partial \vec{\mu}_{\nu}}{\partial Q_{\nu}}\|} 
\frac{R_{\nu}^{LL}}{\|R_{\nu}\|}
\end{equation}
with the label $L$ designating the direction of the near field at the location of the molecule.
To compute $\langle \eta_{\mathrm{pol}} \rangle)$ (see Table~\ref{tab:DFTparam}) 
we numerically average $\eta_{\mathrm{pol}}$ over all possible orientations of the molecule, 
while keeping the IR and optical local field collinear. 

\subsection{Orientation and number of molecules contributing to the IR/optical process}
From the DFT calculations we compute the molecular parameters 
for several cases of interest and report their values in Table \ref{tab:DFTparam}. 
Two orientations (main axis of the molecule parallel to both local fields or fully random molecular orientation) are considered. 
Two options are also considered for the coverage: one monolayer covering the planar parts of the 
metallic nanostructure or a superposition of layers filling the entire volume 
where the fields are localized. We use the IR/optical mode volumes 
$V_\mathrm{IR/opt}$ (given below), the molar mass ($M=0.1102$ kg/mol), volume density ($\rho=1077$ kg/m$^3$) 
or surface density ($\rho_{\mathrm{S}}=6.8\cdot10^{18}$ m$^{-2}$) of thiophenol 
to estimate the number $N_{\textsc{ir}}$ ($N_{\mathrm{opt}}$) of molecules participating in the IR (optical) process.
\\

\section{Numerical simulation of the antenna's optical response}\label{sec:antenna}
\begin{figure*}[t!]
	\xincludegraphics[width=0.5\columnwidth,label=\hspace{-6pt}\textbf{(a)},fontsize=\large]
	{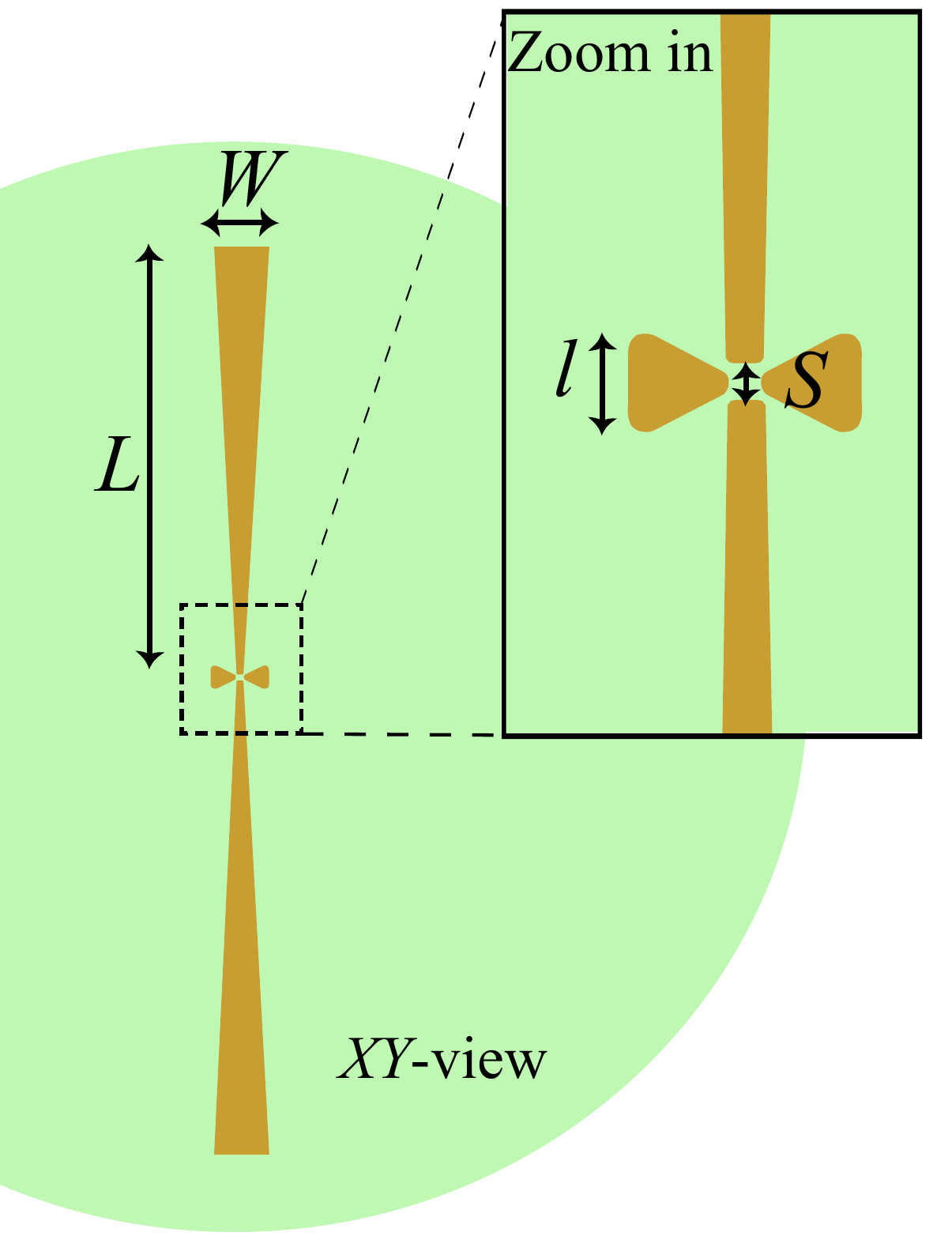}
	\hfill
	\xincludegraphics[width=0.7\columnwidth,label=\hspace{-6pt}\textbf{(b)},fontsize=\large]{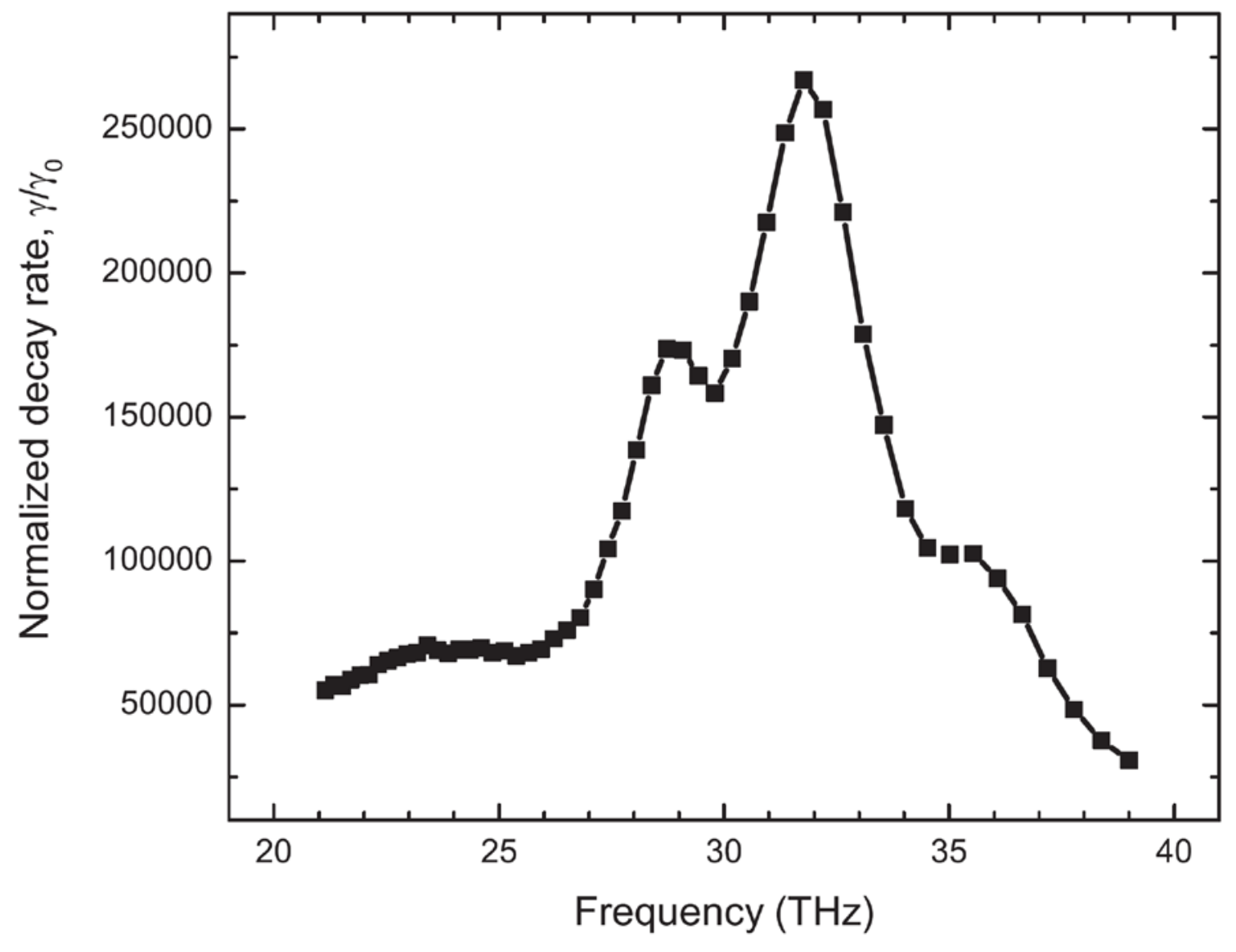}
	\hfill
	\xincludegraphics[width=0.68\columnwidth,label=\hspace{-18pt} \textbf{(c)},fontsize=\large]{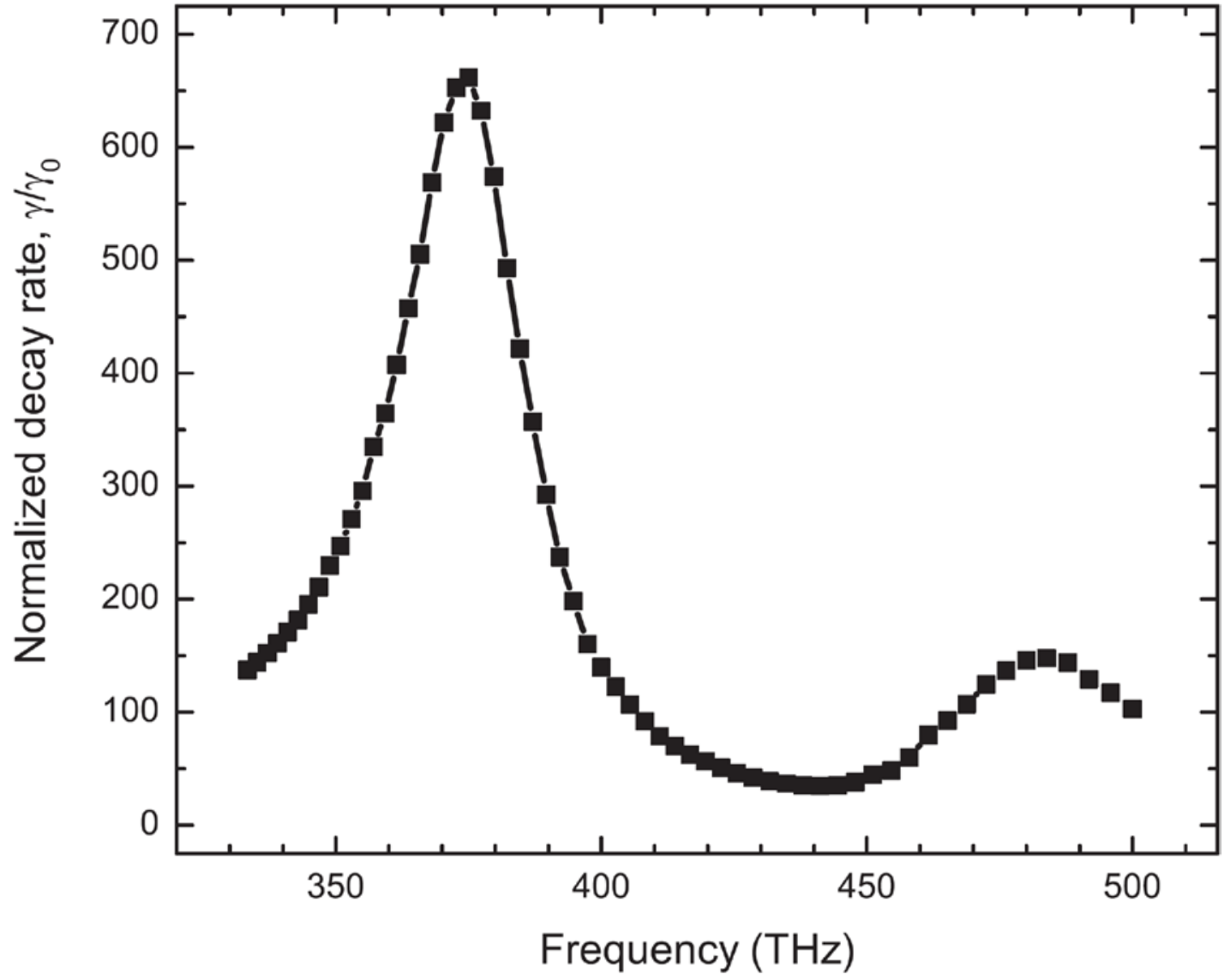}
	\vspace{-5pt}
	\caption{
	\textbf{(a)} Geometry of the dual antenna considered in our manuscript. 
	The parameters $W=70$ nm, $L=2.25~\mu$m, $l=90$ nm, $S=25$ nm 
	are used to run the numerical simulations and evaluate the performance of our conversion scheme. 
	The antenna edges are rounded with 4 and 1.3 nm radii of curvature for the optical and IR antenna, respectively.	
	\textbf{(b-c)} Normalized decay rate of an emitter placed at the center of the dual antenna 
	in the mid-IR \textbf{(b)} and in the visible range \textbf{(c)}.
	}
	\label{Fig:antenna}	
	\vspace{-10pt}
\end{figure*} 
%\subsection{Design of the dual antenna}
Our dual antenna consists of two gold bowtie structures. We set 
the gap between the tips of both antennas ($S=25$ nm) so that the  design 
can be fabricated using current nanofabrication techniques such as focused ion beam milling or advanced \emph{e}-beam lithography.
We select the other structural parameters (Fig. \ref{Fig:antenna}) in order to obtain 
appropriate resonances both in the mid-IR (length $L$ and width $W$) and in the optical domain (short length $l$). 
In our design the $24$ nm high nanostructure lies on top of 
a gold substrate. They are separated by an inactive dielectric layer ($n=1.47$) of $8\ \mu$m thickness.
This substrate reflects the incoming IR field and creates an interference pattern that improves IR absorption as shown in a previous study 
\cite{brown_fan-shaped_2015}. 
\\

\subsection{Numerical calculations}
We use a 3D FEM software (\textsc{Comsol Multiphysics}) to evaluate our dual-antenna design. 
A Drude-Lorentz model describes the electromagnetic response of gold fitted from experimental data \cite{lide_crc_2006}. 
For the calculation in the optical range a dielectric layer (ITO) 
with refractive index $n=1.94$ and thickness $52$~nm  is added below the antenna \cite{sundaramurthy_field_2005}.

A dipolar emitter is placed in the center of our structure to evaluate the local density 
of electromagnetic states inside the antenna. Figure \ref{Fig:antenna} shows the modification of the radiated power 
as a function of the oscillation frequency of the dipole. Based on these plots, we model the response 
of the structure to an incoming optical field and to an incoming mid-IR field (around $32$~THz) as being each dominated by  a single resonance with Lorentzian profile. 
We thus use a multi-Lorentzian fit to extract the relevant linewidths and total decay rates. 
Through the Purcell formula \cite{schmidt_linking_2017}, 
we could estimate the corresponding effective mode volumes. 
Additional integrals are computed to determine the losses originating 
from absorption in the metal and determine the ratio between intrinsic and radiative losses at the resonance frequencies.
All parameters are shown in Table \ref{tab:param}.

\begin{table}[h!]
\setlength\extrarowheight{4pt}
\begin{tabular}{|c|c|c|}
  \hline
  Parameter & Mid-IR & VIS/NIR \tabularnewline 
  \hline
  $\kappa^{{\textsc{ir}}/\mathrm{opt}}/2\pi$ [THz] & $3.2$ & $26.7$ \tabularnewline 
  \hline
  $\eta_{{\textsc{ir}}/\mathrm{opt}}$ & $0.52$ & $0.73$ \tabularnewline 
  \hline
  $V_{{\textsc{ir}}/\mathrm{opt}}$ [m$^{3}$]& $2.6\cdot10^{-21}$ & $1.0\cdot10^{-21}$ \tabularnewline
  \hline  
\end{tabular}
\caption{Antenna parameters as obtained from our FEM simulations. $\kappa^{{\textsc{ir}}/\mathrm{opt}}$ are the total decay rates of IR, respectively optical, energy stored in the antenna,  $\eta_{{\textsc{ir}}/\mathrm{opt}}$ are the respective radiative efficiencies defined as $\kappa_\mathrm{ex}^{{\textsc{ir}}/\mathrm{opt}}/ \kappa^{{\textsc{ir}}/\mathrm{opt}}$, and $V_{{\textsc{ir}}/\mathrm{opt}}$ are the mode volumes.}
\label{tab:param}
\end{table}

\subsection{Spatial overlap - $\eta_{\mathrm{mode}}$}
The spatial overlap between the two electromagnetic modes is computed numerically 
from the field distributions of both modes according to 
\begin{equation}
\eta_{\mathrm{mode}}=
\frac{\left({\displaystyle\int}\left|\vec{E}_{\textsc{ir}}\cdot
\vec{E}_{\mathrm{opt}}\right|\mathrm{d}A\right)^2}
{\left({\displaystyle\int}\left|\vec{E}_{\textsc{ir}}\cdot
\vec{E}_{\textsc{ir}}\right|\mathrm{d}A\right)
\left({\displaystyle\int}\left|\vec{E}_{\mathrm{opt}}\cdot
\vec{E}_{\mathrm{opt}}\right|\mathrm{d}A\right)}
\end{equation}
Imperfect overlap can result from polarization and/or confinement mismatch. 
In our case the dominant mismatch is that between the spatial extents of the two modes. 
In the regions where both fields are confined their polarization mismatch is on the contrary negligible.

\section{Linear array of converters}\label{sec:array}

We discuss the different contributions to the optical noise starting from the expression for 
$\bar{n}_f$, Eq. (\ref{eq:popnu}) in the main text. 
When $\Gamma_{\mathrm{opt}} \ll \Gamma_{\nu}^*$ the equation for the vibrational population 
splits into three different factors identified as thermal $\bar{n}_{\mathrm{th}}$, 
dynamical backaction $\bar{n}_{\mathrm{dba}}$ and quantum backaction noises $\bar{n}_{\mathrm{qba}}$, respectively:
\begin{equation}
\bar{n}_f\simeq \bar{n}_{\mathrm{th}}-\frac{\Gamma_{\mathrm{opt}}}{\Gamma_{\nu}^*}\bar{n}_{\mathrm{th}}
+\frac{A^+}{\Gamma_{\nu}^*}\left(1-\frac{\Gamma_{\mathrm{opt}}}{\Gamma_{\nu}^*}\right).
\label{eq:popvib}
\end{equation}

For sensing applications it is enlightening to study how the contributions 
from the different noise terms are affected when considering an array of converters 
coherently illuminated by the IR field and the pump laser. 
We describe a linear array of $N_\mathrm{conv}$ optomechanical converters. 
For simplicity we consider identical converters 
separated uniformly with a spacing 
$d<\lambda_{\mathrm{opt}}<\lambda_{\textsc{ir}}$ 
in order to avoid multiple maxima in the radiation pattern of the array. 
If all converters are excited in phase, the described configuration is known 
as the broadside configuration and the maximum radiation is directed 
normal to the array axis. 
We assume that the  optical pump power is split among the 
antennas\footnote{The diffraction limit for both beams being largely different, 
we note that multiple converters fit under a focused IR spot. 
In that case the IR power per converter would not scale down.}, 
so that the pump power per antenna is diluted according to 
$|\alpha^{(i)}|^{2}=\frac{1}{N_\mathrm{conv}}|\alpha^{(0)}|^{2}$ 
and the optomechanical coupling rate scales as 
$\frac{1}{N_\mathrm{conv}}$, i.e. $A^{\pm\ (i)}=\frac{1}{N_\mathrm{conv}} A^{\pm\ (0)}$. 
In this section, we use the superscripts $(0)$ vs. $(i)$ to designate quantities computed under single-converter operation vs. array operation (both of them refer to single-converter quantities).

\paragraph{Thermal regime :}
If the backaction effects are weak at a single-converter level 
$\Gamma_{\mathrm{opt}}^{(0)}<\Gamma_{\nu}^{*,(0)}$, 
the power dilution leads to $\Gamma_{\mathrm{opt}}^{(i)}\ll \Gamma_{\nu}^{*,(i)}$. 
The expression of the final population $\bar{n}_f$ (Eq. \ref{eq:popvib}) 
shows that in this case thermal noise is the main contribution to the total noise.

In the far field, constructive interference among the fields emitted 
from individual antennas sharpens the pattern of coherent radiation \cite{dregely_3d_2011} 
so that the total IR converted signal in this direction scales as \cite{balanis_antenna_2005} 
$S^{\mathrm{out},(N_\mathrm{conv})}_{\textsc{ir}\mathrm{\rightarrow opt}}=\mathrm{(array\ factor)}^2\cdot S^{\mathrm{out},(i)}_{\textsc{ir}\mathrm{\rightarrow opt}}$ 
which results in $S^{\mathrm{out},(N_\mathrm{conv})}_{\textsc{ir}\mathrm{\rightarrow opt}}\simeq N_\mathrm{conv}^2 S^{\mathrm{out},(i)}_{\textsc{ir}\mathrm{\rightarrow opt}}$ 
along the direction of maximum radiation for a broadside array. 
On the contrary if the converters are sufficiently spaced to avoid 
any near-field coupling the thermal emission remains incoherent 
and quasi-isotropic. 

We combine the factors related to the power dilution and 
to the directivity of the linear array to describe the SNR of the array 
in the regime dominated by thermal noise: 
\begin{equation}
\mathrm{SNR}^{(N_\mathrm{conv})}\simeq N_\mathrm{conv}^2\cdot\frac{S^{\mathrm{out},(i)}_{\textsc{ir}\mathrm{\rightarrow opt}}}{S^{\mathrm{out},(i)}_{\mathrm{th}}}=
N_\mathrm{conv}^2\cdot\frac{1}{N_\mathrm{conv}}\cdot\frac{S^{\mathrm{out},(0)}_{\textsc{ir}\mathrm{\rightarrow opt}}}{S^{\mathrm{out},(0)}_{\mathrm{th}}}.
\end{equation}

\paragraph{Zero-temperature limit :}
In the case where the thermal population of the vibrational mode 
is negligible ($\bar{n}_{\mathrm{th}}\sim 0$), corresponding to backaction noise dominating over thermal noise,
the incoming power dilution lowers equivalently the converted signal 
and the output noise per antenna,
so that the SNR of a sufficiently large array in this regime is given by~:
\begin{equation}
\mathrm{SNR}^{(N_\mathrm{conv})}\simeq N_\mathrm{conv}^2\cdot\frac{S^{\mathrm{out},(i)}_{\textsc{ir}\mathrm{\rightarrow opt}}}{S^{\mathrm{out},(i)}_{\mathrm{ba}}}=
N_\mathrm{conv}^2 \cdot 1 \cdot \frac{S^{\mathrm{out},(0)}_{\textsc{ir}\mathrm{\rightarrow opt}}}{S^{\mathrm{out},(0)}_{\mathrm{ba}}}.
\end{equation}

\bibliography{../Bib/THzLibraryV5}

\end{document}